\documentclass[11pt,a4paper,preprintnumbers,amsmath,amssymb,nofootinbib]{article}
\pdfoutput=1
\usepackage{feynmp-auto,expdlist}
\usepackage{amsmath,amsfonts,amssymb}
\usepackage{graphicx}
\usepackage{enumerate}
\usepackage{hyperref}
\usepackage{axodraw2}
\usepackage{latexsym}
\usepackage{hepnicenames}
\usepackage{enumerate}
\usepackage{soul}
\usepackage[normalem]{ulem}
\usepackage{wasysym}
\usepackage{makecell}
\usepackage{bbm}

\usepackage{tikz} 
\usepackage{tikz-feynman} 
\usepackage{tikz-feynhand} 
\tikzfeynmanset{compat=1.1.0} 

\font\tenrsfs=rsfs10 at 12pt
\font\sevenrsfs=rsfs7
\font\fiversfs=rsfs5
\newfam\rsfsfam
\textfont\rsfsfam=\tenrsfs
\scriptfont\rsfsfam=\sevenrsfs
\scriptscriptfont\rsfsfam=\fiversfs

\numberwithin{equation}{section}

\usepackage{braket}
\usepackage{titling}
\usepackage{amsmath}
\usepackage{slashed}
\usepackage{amssymb}
\usepackage{epsfig}
\usepackage{graphicx}
\usepackage{color,xcolor}
\usepackage{rotating}
\usepackage{hyperref}
\usepackage[margin=1.in]{geometry}
\usepackage[compress,numbers,sort]{natbib}
\usepackage{colortbl}
\usepackage{pdflscape}
\usepackage{mathtools}
\usepackage{comment}
\definecolor{Gray}{gray}{0.95}
\definecolor{RGray}{gray}{0.85}
\definecolor{CGray}{gray}{0.93}
\definecolor{piggypink}{rgb}{0.99, 0.87, 0.9}
\definecolor{babyblue}{rgb}{.67,.83,.99}

\newcommand{\SU}{{\rm SU}}

\newcommand{\U}{{\rm U}}

\newcommand{\M}{{\cal M}}
\renewcommand{\L}{{\cal L}}

\newcommand{\SM}{{\rm SM}}
\newcommand{\hc}{{\rm h.c.}}
\newcommand{\eff}{{\rm eff}}

\newcommand{\DM}{{\rm DM}}
\newcommand{\RH}{{\rm RH}}
\newcommand{\cm}{{\rm cm}}
\newcommand{\GeV}{{\rm GeV}}
\newcommand{\TeV}{{\rm TeV}}
\newcommand{\MeV}{{\rm MeV}}
\newcommand{\keV}{{\rm keV}}

\newcommand{\1}{{\textbf{1}}}

\newcommand{\3}{{\textbf{3}}}

\newcommand{\la}{\lambda}
\newcommand{\eps}{\epsilon}

\newcommand{\ol}[1]{\overline{#1}}
\newcommand{\wt}[1]{\widetilde{#1}}

\newcommand{\lrpartial}[1]{\overset{\leftrightarrow}{\partial_{#1}}}
\newcommand{\vmol}{v_{\rm M\o l}}

\definecolor{nicered}{rgb}{0.7,0.1,0.1}
\definecolor{nicegreen}{rgb}{0.1,0.5,0.1}
\definecolor{red}{rgb}{1.0, 0, 0}
\definecolor{niceblue}{rgb}{0,0,0.8}
\definecolor{red}{rgb}{1.0, 0, 0}
\hypersetup{colorlinks,bookmarksopen,bookmarksnumbered,
linkcolor=blus,pdfstartview=FitH,urlcolor=rossos,citecolor=verde}
\allowdisplaybreaks
\hypersetup{colorlinks,citecolor= nicegreen,linkcolor= niceblue,urlcolor=nicered}

\definecolor{rosso}{cmyk}{0,1,1,0.4}
\definecolor{rossos}{cmyk}{0,1,1,0.55}
\definecolor{rossoc}{cmyk}{0,1,1,0.2}
\definecolor{blu}{cmyk}{1,1,0,0.3}
\definecolor{blus}{cmyk}{1,1,0,0.6}
\definecolor{bluc}{cmyk}{1,1,0,0.1}
\definecolor{verde}{cmyk}{0.92,0,0.59,0.25}
\definecolor{verdec}{cmyk}{0.92,0,0.59,0.15}
\definecolor{verdes}{cmyk}{0.92,0,0.59,0.4}

\def\eq#1{{Eq.~(\ref{#1})}}

\def\vev#1{\left\langle #1\right\rangle}

\def\ol#1{\overline{#1}}

\renewcommand{\d}{{\rm d}}
\renewcommand{\bar}{\overline}

\newcommand{\beq}{\begin{equation}}
\newcommand{\eeq}{\end{equation}}
\newcommand{\bea}{\begin{eqnarray}}
\newcommand{\eea}{\end{eqnarray}}

\renewcommand{\[}{\left[}
\renewcommand{\]}{\right]}
\renewcommand{\(}{\left(}
\renewcommand{\)}{\right)}

\renewcommand{\arraystretch}{1.3}

\graphicspath{{figs/}} 

\begin{document}
\begin{titlepage}
\setcounter{page}{0} 

\begin{flushright}
KEK-TH-2649
\end{flushright}

\begin{center}

\vskip .55in

\begingroup
\centering
\Large\bf 
Multi-Component Dark Matter from Minimal Flavor Violation

\endgroup

\vskip .4in

\renewcommand{\thefootnote}{\fnsymbol{footnote}}
{
Federico Mescia$^{(a)}$\footnote{
  \href{mailto:federico.mescia@lnf.infn.it}
  {\tt federico.mescia@lnf.infn.it}}\footnote[6]{On leave of absence from Universitat de Barcelona},
Shohei Okawa$^{(b,c)}$\footnote{
  \href{mailto:shohei.okawa@kek.jp}
  {\tt shohei.okawa@kek.jp}},
Keyun Wu$^{(c)}$\footnote{
  \href{mailto:keyunwu@fqa.ub.edu}
  {\tt keyunwu@fqa.ub.edu}}
}

\vskip 0.4in

\begingroup\small
\begin{minipage}[t]{1.0\textwidth}
\centering\renewcommand{\arraystretch}{1.0}
{\it
\begin{tabular}{c@{\,}l}
$^{(a)}$
& Istituto Nazionale di Fisica Nucleare (INFN), Laboratori Nazionali di Frascati,\\
& Via E. Fermi 54, C.P.~13, I-00044 Frascati, Italy\\[2mm]
$^{(b)}$
& KEK Theory Center, Tsukuba, Ibaraki 305–0801, Japan \\
$^{(c)}$
& Departament de F\'{i}sica Qu\`{a}ntica i Astrof\'{i}sica, Institut de Ci\`{e}ncies del Cosmos (ICCUB), \\
& Universitat de Barcelona, Mart\'{i} i Franqu\`{e}s 1, E-08028 Barcelona, Spain \\[2mm]
\end{tabular}
}
\end{minipage}
\endgroup

\end{center}

\vskip .4in

\begin{abstract}\noindent
Minimal Flavor Violation (MFV) offers an appealing framework for exploring physics beyond the Standard Model. 
Interestingly, within the MFV framework, a new colorless field that transforms non-trivially under a global $\SU(3)^3$ quark flavor group can naturally be stable. 
Such a new field is thus a promising dark matter candidate, provided it is electrically neutral. 
We extend the MFV framework for dark matter and demonstrate that dark matter can naturally be multi-component across a broad parameter space. 
For illustration, we consider a gauge singlet, flavor triplet scalar field and identify parameter spaces for multi-component dark matter, 
where only the lightest flavor component is absolutely stable and heavy flavor components are decaying with lifetimes sufficiently longer than the age of the universe. 
Phenomenological, cosmological and astrophysical aspects of multi-component flavored dark matter are briefly discussed. 
\end{abstract}
\end{titlepage}

\thispagestyle{empty}
\renewcommand{\thefootnote}{\#\arabic{footnote}}
\setcounter{footnote}{0}

\tableofcontents

\clearpage
\setcounter{page}{1}

\section{Introduction}
\label{sec:introduction}

Matter content of the Standard Model (SM) comprises five different gauge representations of Weyl fermions, called quarks and leptons. 
In each representation, there exist three species, or flavors. 
The SM gauge interactions do not distinguish these three fermion flavors in the same representation, leading to a global $\U(3)^5$ flavor symmetry in the gauge sector. 
This flavor symmetry is explicitly broken by quark and lepton Yukawa interactions to the Higgs doublet field. 
In particular, the breaking of the $\SU(3)^5$ subgroup of $\U(3)^5$ governs mixing patterns among different flavors, thereby introducing non-trivial flavor violating processes at low energies. 
Flavor violation in the quark sector is characterized by the hierarchical quark masses and the Cabibbo-Kobayashi-Maskawa (CKM) mixing matrix, whose unique mixing pattern has been confirmed experimentally with a good accuracy. 
The lepton sector does not exhibit any flavor mixing due to the absence of neutrino masses in the SM. 
The global $\SU(3)_{\ell_R} \times \SU(3)_{e_R}$ lepton flavor symmetry is broken down to $\U(1)_{L_e-L_\mu} \times \U(1)_{L_\mu-L_\tau}$ by the lepton masses.

New interactions from physics beyond the SM can generally provide independent sources of flavor violation. 
The resulting modifications to flavor violating observables are faced with current precise measurements, if one expects new particles mediating the flavor violation to reside around TeV scales, as motivated by the naturalness problem. 
This strong flavor constraints can be circumvented by invoking the Minimal Flavor Violation (MFV) hypothesis~\cite{Chivukula:1987py, Hall:1990ac, Buras:2000dm, DAmbrosio:2002vsn}, which dictates that new physics interactions also respect the $\U(3)^5$ flavor symmetry with the only breaking sources stemming from the quark and lepton Yukawa matrices, $Y_{u,d,e}$. 
Formally, the MFV interaction structure can be achieved by promoting the Yukawa matrices to spurious fields transforming under the flavor group, 
$\U(3)^5=\U(3)_{q_L}\times\U(3)_{u_R}\times\U(3)_{d_R}\times\U(3)_{\ell_L}\times\U(3)_{e_R}$:
\beq
Y_u \sim (\3,\bar{\3},\1,\1,\1) \,,\quad
Y_d \sim (\3,\1,\bar{\3},\1,\1) \,,\quad
Y_e \sim (\1,\1,\1,\3,\bar{\3}) \,.
\eeq
This transformation rule assigned to the Yukawa matrices assures the (apparent) flavor invariance of the SM Yukawa interaction Lagrangian, 
\beq
\L_{\rm yuk} = - \bar{q}_L Y_u \wt{H} u_R 
                - \bar{q}_L Y_d H d_R 
                - \bar{\ell}_L Y_e H e_R + \hc \,,
\eeq
with $\wt{H}=i\sigma_2 H^*$. 
Implementing the MFV structure in new physics models is straightforward. 
For a new interaction operator ${\cal O}_{ij\dots}$ ($i,j,\dots$ denote flavor indices), its coupling $C_{ij\dots}$ is parameterized by a series of spurion insertions so that the corresponding interaction is invariant under the flavor transformation. 
For example, when we consider a flavor violating operator ${\cal O}_{ij} = (\bar{u}_{Ri} \gamma^\mu u_{Rj})(X^\dag i \lrpartial{\mu} X)$ with a gauge and flavor singlet scalar $X$, the MFV requires $C_{ij}$ to take the form, 
\beq
C_{ij} = c_0\,\delta_{ij} + \eps \, c_1 (Y_u^\dagger Y_u)_{ij} 
    + \eps^2 \[c_2 (Y_u^\dag Y_u Y_u^\dag Y_u)_{ij} + c_2^\prime (Y_u^\dag Y_d Y_d^\dag Y_u)_{ij}\] 
    + \dots \,,
\eeq
where the ellipsis denotes further spurion insertions. 
Flavor violating effects from this new interaction are suppressed by the power of the quark Yukawa couplings, the CKM off-diagonal elements and a potentially small MFV expansion parameter $\eps$.\footnote{Here we implicitly assume $\eps\ll1$ and the MFV structure is linearly realized. The MFV implementation can be generalized to $\eps\sim{\cal O}(1)$ case in which all spurion insertions are equally contributing and have to be resummed appropriately, rendering the MFV non-linearly realized \cite{Kagan:2009bn}.}

Remarkably, the MFV in new physics models can guarantee the stability of dark matter (DM). 
It is shown in \cite{Batell:2011tc} that within the MFV framework, the lightest state of a new colorless field $\chi$ that transforms under the quark flavor subgroup, i.e. ${\cal G}_F=\SU(3)_{q_L} \times \SU(3)_{u_R} \times \SU(3)_{d_R}$, is absolutely stable, even if including all higher dimensional operators. 
That lightest particle is therefore an excellent DM candidate if electrically neutral. 
This stability discussion relies only on the invariance under the color and flavor groups within the MFV and does not depend on spin and $\SU(2)_L\times\U(1)_Y$ representation of $\chi$. 
In \cite{Batell:2011tc}, they focus on a gauge singlet scalar DM case and study cosmological and phenomenological implications, ranging from the conventional freeze-out production to characteristic collider signals as well as effects on flavor changing neutral current (FCNC) processes. 
The study of \cite{Batell:2011tc} was followed up in detail by \cite{Lopez-Honorez:2013wla}, where they surveyed the traditional Weakly Interacting Massive Particle (WIMP) DM parameter space for the simplest singlet scalar. 

In this paper, we demonstrate that within the MFV framework, 
DM can naturally be multi-component in certain parameter spaces. 
Although this possibility was very briefly mentioned in \cite{Batell:2011tc}, no follow-up studies in this direction have been published thus far. 
We focus on a model featuring an $\SU(2)_L\times\U(1)_Y$ singlet flavored scalar field, and evaluate lifetimes of heavy flavor components. 
We then identify parameter spaces where more than one flavor component has sufficient longevity to serve as DM. 
Our work is supplemented by studying DM production and direct detection bounds and by giving general comments on phenomenological and cosmological  aspects of multi-component flavored DM scenarios. 

This paper is organized as follows. 
We review in Section \ref{sec:MFVDM} the formulation of the DM stabilization in new physics models based on the MFV principle. 
An example model we focus on in this paper is explicitly introduced in Section \ref{sec:model}. 
Then, in Section \ref{sec:decay}, we evaluate decay widths of heavier states into lighter states and identify parameter spaces where DM can comprise multiple states. 
In Sections \ref{sec:results} and \ref{sec:discussion}, 
we present our main results and discuss the outlook for phenomenological works in future. 
In appendices, we provide calculation tools to study multi-body decays and DM production and direct detection.

\section{Stability of flavored dark matter}
\label{sec:MFVDM}

To formulate the DM stability under the MFV, 
let $\chi$ be a singlet of $\SU(3)_c$ and a multiplet of ${\cal G}_F=\SU(3)_{q_L} \times \SU(3)_{u_R} \times \SU(3)_{d_R}$. 
The representation of $\chi$ under ${\cal G}_F$ is specified by the Dynkin coefficients $(n_i, m_i)$ of the corresponding $\SU(3)_i$ flavor groups: 
\beq
\chi \sim (n_{q_L}, m_{q_L}) \times (n_{u_R}, m_{u_R}) \times (n_{d_R}, m_{d_R}) \,,
\eeq
where we do not specify the spin and $\SU(2)_L \times \U(1)_Y$ representation of $\chi$, which are irrelevant to the stability discussion. 
General decay vertices of $\chi$ into SM fields formally take the form, 
\begin{align}
{\cal O}_{\rm decay} 
    & = \chi \times 
        \underbrace{q_L\dots}_{A} \, 
        \underbrace{\bar{q}_L\dots}_{\bar{A}} \, 
        \underbrace{u_R\dots}_{B} \, 
        \underbrace{\bar{u}_R\dots}_{\bar{B}} \, 
        \underbrace{d_R\dots}_{C} \, 
        \underbrace{\bar{d}_R\dots}_{\bar{C}} \,
        \underbrace{Y_u\dots}_{D} \, 
        \underbrace{Y_u^\dag\dots}_{\bar{D}} \, 
        \underbrace{Y_d\dots}_{E} \, 
        \underbrace{Y_d^\dag\dots}_{\bar{E}} 
        \times {\cal O}_{\rm weak} \,,
\end{align}
where ${\cal O}_{\rm weak}$ denotes a potential weak operator having no color nor flavor to make ${\cal O}_{\rm decay}$ invariant under the Lorentz and $\SU(2)_L\times\U(1)_Y$ transformation. 
The color and quark flavor invariance of ${\cal O}_{\rm decay}$ requires that the triality of each SU(3) group vanishes, i.e.
\begin{align}
&(A+B+C-\bar{A}-\bar{B}-\bar{C})\,{\rm mod}\,3=0\,,\\
&(n_{q_L}-m_{q_L}+A-\bar{A}+D-\bar{D}+E-\bar{E})\,{\rm mod}\,3=0\,,\\
&(n_{u_R}-m_{u_R}+B-\bar{B}-D+\bar{D})\,{\rm mod}\,3=0\,,\\
&(n_{d_R}-m_{d_R}+C-\bar{C}-E+\bar{E})\,{\rm mod}\,3=0\,,
\end{align}
which in turn requires the {\it flavor triality} of $\chi$ to vanish: 
$(n_\chi - m_\chi)\,{\rm mod}\,3=0$ where $n_\chi=n_{q_L}+n_{u_R}+n_{d_R}$ and $m_\chi=m_{q_L}+m_{u_R}+m_{d_R}$. 
In other words, if we choose a flavor representation for $\chi$ such that the flavor triality is non-vanishing, i.e. 
\beq
(n_\chi - m_\chi)\,{\rm mod}\,3\neq0 \,,
\label{eq:triality}
\eeq
then ${\cal O}_{\rm decay}$ is forbidden and $\chi$ is absolutely stable \cite{Batell:2011tc}. 
If the lightest state of $\chi$ is neutral, it is a good DM candidate. 
It should be noted that we did not restrict the mass dimension of ${\cal O}_{\rm decay}$ and hence this stability discussion can apply for all higher dimensional operators. 

After that pioneering work \cite{Batell:2011tc}, 
various flavored DM models have been studied. 
In \cite{Lopez-Honorez:2013wla}, they scan the conventional WIMP regime of the simplest flavored scalar DM, originally proposed in \cite{Batell:2011tc}, and evaluate the impact of Higgs portal couplings to the DM phenomenology. 
In \cite{Batell:2013zwa}, they find out general features of supersymmetric flavored DM models as well as provide a deeper insight into the role of the flavor symmetries in the DM stability. 
In that paper, it is emphasized that the MFV is sufficient but not necessary for the DM stability, and the most essential is the flavor triality condition Eq.~(\ref{eq:triality}). 
In fact, \cite{Bishara:2015mha} shows that in a class of new physics models, even though the MFV is not respected, a new flavored state can have the absolute stability as long as the flavor triality condition is fulfilled. 
See also \cite{Bishara:2014gwa}, where the MFV is crucial for accidental longevity of asymmetric DM. 
The concept of flavored DM was extended to incorporate a dark flavor symmetry $\SU(3)_\chi$ under which a DM field is charged while all SM fields are not \cite{Agrawal:2014aoa}. 
This extended framework abandons the MFV principle and necessitates an additional global symmetry imposed by hand to guarantee the DM stability, but predicts richer flavor phenomenology. 
(See \cite{Kile:2011mn, Kamenik:2011nb, Agrawal:2011ze} for related studies on DM carrying a flavor charge, where the DM stability is not necessarily attributed to the MFV.)
Currently, the terminology of flavored DM mainly points to the extended framework, but in this paper we build on the original MFV framework in \cite{Batell:2011tc}. 

In general, heavy states of $\chi$ can decay into lighter ones. 
If all heavy states decay quickly, only the lightest one is stable and DM. 
The simplest case of such a single component flavored DM is studied in \cite{Batell:2011tc, Lopez-Honorez:2013wla}. 
On the other hand, some heavy states can constitute part of cosmological DM if long-lived enough. 
Lifetimes of heavy states will depend on several factors, such as mass splitting with the lightest state, interaction operators triggering decay, and cutoff scales if heavy states are decaying mostly due to higher dimensional operators. 
In the following sections, we will take an example model and show that more than one component of a flavored new field can be stable and constitute a significant portion of DM in the universe.

\section{Model}
\label{sec:model}

We consider a gauge singlet scalar field $S$, which transforms like $(\1,\3,\1)$ under the quark flavor group ${\cal G}_F$. 
The choice of the flavor representation for $S$ is different from the one studied in \cite{Batell:2011tc, Lopez-Honorez:2013wla}, but it is irrelevant to our main conclusion. 
Under the MFV hypothesis, all mass and interaction terms respect the ${\cal G}_F$ symmetry with the only breaking sources from the quark Yukawa matrices. 
The general interaction Lagrangian takes the form
\beq
\L = \L_\SM + (\partial_\mu S_i^*) (\partial_\mu S_i) - V(H,S) + \L_{d > 4} \,,
\eeq
where $i=1,2,3$ is the flavor index and $\L_{d > 4}$ denotes higher dimensional operators composed of SM fields and $S$. 
In this section, we provide renormalizable interactions of the flavored scalar field $S$ and a set of dimension-6 operators involving two flavored scalar fields, which induce decays of heavy flavor components.

\subsection{Renormalizable interactions}

The scalar potential takes the form, 
\begin{align}
V(H,S) 
    & = m_S^2 \, S_i^* \( a_0\,\delta_{ij} + \eps \, a_1 (Y_u^\dag Y_u)_{ij} + \dots \) S_j \nonumber\\
    & \quad + \la \, S_i^*\( b_0\,\delta_{ij} + \eps \, b_1 (Y_u^\dag Y_u)_{ij} + \dots \) S_j (H^\dag H) \nonumber\\
    & \quad + \( \la_0\,\delta_{ij} \delta_{kl} + \eps\,\la_1\delta_{ij} (Y_u^\dag Y_u)_{kl} + \dots \) \, S_i^* S_j S_k^* S_l \,,
\label{eq:VS}
\end{align}
where the flavor indices run over $i,j=1,2,3$, 
$\la_i$ are all real parameters, 
$\eps$ is a small MFV expansion parameter, 
$a_0, a_1, b_0, b_1$ are ${\cal O}(1)$ coefficients, and 
the ellipsis indicates further MFV spurion insertions involving four or more Yukawa matrices, 
which we neglect here. 
Effects of those higher order terms will be discussed in Section \ref{sec:NNLO}.

Without loss of generality, the up-quark Yukawa matrix is expressed as 
$(Y_u)_{ij} = (V^\dag \hat{Y}_u)_{ij}$ with $(\hat{Y}_u)_{ij} = y_u^i \, \delta_{ij}$ and $V$ the CKM matrix. 
After the electroweak (EW) symmetry breaking, 
the physical masses and Higgs portal couplings of $S_i$ are expressed by 
\beq
V(H,S) \supset M_i^2 (S_i^* S_i) + \frac{\la_{hSi}}{2} (2vh+h^2) (S_i^* S_i) \,,
\label{eq:V_EWSB}
\eeq
where $v=246\,\GeV$ and 
\begin{align}
M_i^2 & = m_i^2 + \frac{\la_{hSi} v^2}{2} \,,\\
m_i^2 & = m_S^2 \( a_0 + \eps\, a_1 (y_u^i)^2 \) \,,\\
\la_{hSi} & = \la \( b_0 + \eps\, b_1 (y_u^i)^2 \) \,.
\end{align}
The mass square difference of the flavored scalars is determined by the up-quark Yukawa couplings,
\beq
M_j^2-M_i^2 = \epsilon \(a_1 m_S^2+b_1\frac{\lambda v^2}{2}\)\left\{(y_u^j)^2-(y_u^i)^2\right\} \,.
\label{eq:DelM}
\eeq
It follows from this equation that the ratio of the mass square difference is sharply predicted
\beq
\frac{M_3^2-M_1^2}{M_2^2-M_1^2} 
    = \frac{y_t^2-y_u^2}{y_c^2-y_u^2} 
    \simeq \frac{y_t^2}{y_c^2} \,.
\eeq
Since the sign of $a_1,b_1,\lambda$ is arbitrary, the mass ordering of the flavor components $S_i$ is not fixed from the MFV assumption, and the mass spectrum can be either {\it normal} ($M_1 < M_2 < M_3$) or {\it inverted} ($M_3 < M_2 < M_1$).

A notable feature of the scalar potential Eq.~(\ref{eq:VS}) is that there is no flavor off-diagonal interaction for $S_i$, if the MFV expansion is truncated at the order of $\epsilon$. 
All three scalars are thus individually stable at this order. 
This threefold stability is broken to the stability of the lightest flavored scalar once including higher order terms in the MFV expansion, see Section \ref{sec:NNLO} for further details.
However, the first flavor off-diagonal vertices appear in the scalar potential at the order of $\epsilon^2$, and by taking a small $\epsilon$, one can assure sufficient longevity for the heavy scalars to serve as DM. 

Before proceeding, we would like to mention theoretical constraints on the scalar potential. 
We require the potential to have a global minimum at $\vev{H}\neq0$ and $\vev{S_i}=0$, since a non-vanishing vacuum expectation value (VEV) of $S_i$ breaks the flavor symmetry and triggers instability of DM. 
This requirement also implicitly imposes a bounded-from-below condition, 
which is read from the quartic terms in the potential, 
\begin{align}
V|_{\rm quartic} 
	& = \la_H |H|^4 + \la_0 \(|S_1|^2 + |S_2|^2\)^2 + \(\la_0+\la_1 y_t^2\) |S_3|^4 \nonumber\\
		& \quad + \la_{hS1} \, |H|^2 \(|S_1|^2 + |S_2|^2\) + \la_{hS3} \, |H|^2 |S_3|^2 \nonumber\\
		& \quad + \(2\la_0 + \la_1 y_t^2\) \(|S_1|^2 + |S_2|^2\) |S_3|^2 \,,
\end{align}
where small yukawa couplings $y_{u,c}$ are ignored. 
This potential can be written as 
\beq
V|_{\rm quartic} = \sum_{i,j} \Lambda_{ij} X_i X_j \,,
\eeq
where $X_i\equiv\{|H|^2, \, |S_1|^2+|S_2|^2, \,|S_3|^2\}$ and 
\beq
\Lambda_{ij} 
    \equiv 
    \begin{pmatrix} 
	\la_H & \la_{hS1}/2 & \la_{hS3}/2 \\
	\la_{hS1}/2 & \la_0& (\la_0+\la_t)/2 \\
	\la_{hS3}/2 & (\la_0+\la_t)/2 & \la_t 
    \end{pmatrix} \,,
\eeq
with 
\beq
\la_t \equiv \la_0+\la_1 y_t^2 \,.
\eeq
Then, it follows from co-positivity criteria \cite{Kannike:2012pe} that the bounded-from-below condition is fulfilled if and only if the following inequalities are all satisfied: 
\begin{align}
\la_H & > 0 ,\ \la_0 > 0 ,\ \la_t > 0 \,,\label{eq:VFB1}\\ 
\bar{\Lambda}_{12} & := \la_{hS1}/2+\sqrt{\la_H \la_0} > 0\,,\label{eq:VFB2}\\
\bar{\Lambda}_{13} & := \la_{hS3}/2 + \sqrt{\la_H \la_t} > 0 \,,\label{eq:VFB3}\\
\bar{\Lambda}_{23} & := \(\la_0+\la_t\)/2 + \sqrt{\la_0 \la_t} > 0 \,,\label{eq:VFB4}
\end{align}
and 
\beq
\sqrt{\la_H \la_0 \la_t} + \frac{\la_{hS1}}{2} \sqrt{\la_t} + \frac{\la_{hS3}}{2} \sqrt{\la_0} + \frac{\la_0+\la_t}{2} \sqrt{\la_H} + \sqrt{2\bar{\Lambda}_{12}\bar{\Lambda}_{13}\bar{\Lambda}_{23}} > 0 \,.
\label{eq:VFB5}
\eeq
We have confirmed that these conditions are all satisfied in parameter spaces we focus on in this paper.

\subsection{Dimension-6 operators}
\label{sec:dim6}

Three flavored scalars $S_{1,2,3}$ are individually stable with the scalar potential Eq.~(\ref{eq:VS}) unless taking into account higher order terms in the $\epsilon$ expansion. 
However, inclusion of higher dimensional operators causes the heavy scalars to decay into the lighter ones even at the leading order of $\epsilon$.
Of the most relevance are dimension-6 operators involving two quarks and two flavored scalar fields $S_i$, given by 
\beq
\L_{d=6} = \frac{1}{\Lambda^2} \sum_I c_{ijkl}^I {\cal O}_{ijkl}^I \,,
\eeq
where 
\begin{align}
{\cal O}_{ijkl}^1 
	& = (\bar{q}_{Li} \gamma^\mu q_{Lj}) (S_k^* i \lrpartial{\mu} S_l) \,,
~~~~~~
{\cal O}_{ijkl}^2 
	= (\bar{u}_{Ri} \gamma^\mu u_{Rj}) (S_k^* i \lrpartial{\mu} S_l) \,,\nonumber\\
{\cal O}_{ijkl}^3 
	& = (\bar{d}_{Ri} \gamma^\mu d_{Rj}) (S_k^* i \lrpartial{\mu} S_l)\,,
~~~~~~
{\cal O}_{ijkl}^4 
        = (\bar{q}_{Li} \wt{H} u_{Rj}) (S_k^* S_l) \,,\\
{\cal O}_{ijkl}^5 
	& = (\bar{q}_{Li} H d_{Rj}) (S_k^* S_l) \,.\nonumber
\end{align}
The coefficients $c^I_{ijkl}$ are expanded with respect to the quark Yukawa matrices following the MFV, and for example, we have for ${\cal O}^4_{ijkl}$ 
\begin{align}
c^4_{ijkl} 
	& = c_1 (Y_u)_{il} \delta_{kj} + c_2 (Y_u)_{ij}  \delta_{kl} \nonumber\\
	& + \eps \[ c_3 (Y_u Y_u^\dag Y_u)_{ij} \delta_{kl} 
		+ c_4 (Y_u Y_u^\dag Y_u)_{il}  \delta_{kj} 
		+ c_5 (Y_u)_{ij} (Y_u^\dag Y_u)_{kl} 
		+ c_6 (Y_u)_{il} (Y_u^\dag Y_u)_{jl} \] \nonumber\\
	& + {\cal O}(\eps^2) \,.
\end{align}
Here, we expect $\eps \ll 1$ and ignore higher order terms until Section \ref{sec:NNLO}.

Let us focus on the ${\cal O}^4_{ijkl}$ operator at the order of $\epsilon^0$, 
\beq
\L_{d=6} = \frac{1}{\Lambda^2} \[
	c_1 \(\bar{q}_{Li} (V^\dag \hat{Y}_u)_{ij} S_j\) \wt{H} \(S_k^* \delta_{kl} u_{Rl}\) 
	+ c_2 \(\bar{q}_{Li} (V^\dag \hat{Y}_u)_{ij} \wt{H} u_{Rj}\) \(S_k^* \delta_{kl} S_l\) \]
	+ \hc
\label{eq:O4-LO}
\eeq
After the EW symmetry breaking and taking the up-type quark mass basis (i.e. $u_L \to V^\dagger u_L$), this Lagrangian reduces to 
\begin{align}
\L_{d=6} 
	& = \frac{1}{\Lambda^2} \[
		c_1 \, \bar{u}_i (m_i P_R + m_j P_L) u_j  \(S_j^* S_i\) 
            + c_2 \( m_i \bar{u}_i u_i \) \(S_j^* S_j\) \] \,,
\label{eq:O4-EWSB}
\end{align}
where $u_{i,j}$ denotes the up-quark fields in the mass basis. 
It is easy to see that the $c_2$ term does not induce decay of heavy scalars, 
since it only produces flavor diagonal interactions like $(\bar{u}_i u_i) (S^*_j S_j)$. 
In contrast, the $c_1$ interactions cause heavy scalar decays, whose partial decay width scales as 
\beq
\Gamma(S_i \to S_j u_{Li} \bar{u}_{Rj}) 
	\sim \(\frac{c_1}{\Lambda^2}\)^2 \left\{(m_u^i)^2+(m_u^j)^2\right\} \times \int \d\Phi_3 \,,
\eeq
where $\d\Phi_3$ denotes three-body phase space. 
In the case of the normal spectrum, $S_2$ and $S_3$ are unstable and decaying. 
$S_2$ is expected to decay into $S_1$ with a very suppressed rate because of the mass degeneracy between $S_2$ and $S_1$, 
whereas $S_3$ has a moderately large mass splitting and relatively easily decays into $S_1$ or $S_2$. 
Thus, there will be a parameter space where both $S_1$ and $S_2$ are stable on the cosmological time scale, while $S_3$ decays away in the early universe. 
In other case, $S_3$ can also be stable if the mass splitting is small (e.g. by taking $\epsilon \to 0$) or the cutoff scale $\Lambda$ is high enough, resulting in all three components being DM. 
The same argument can apply for the inverted spectrum, and 
some or all of the flavored scalars can form DM, depending on their mass splittings and the magnitude of the cutoff scale. 
In either case, if the heavy scalars are unstable, 
they have to decay prior to $\sim1\,{\rm sec}$ to avoid Big Bang Nucleosynthesis (BBN) bounds, and 
if they are long-lived, their lifetimes must be longer than the age of the universe, $t_U \simeq 13.8\times10^9\,{\rm yrs}$. 
These bounds on the heavy scalar lifetimes constrain the cutoff scale $\Lambda$.

\section{Decay of heavy states}
\label{sec:decay}

In this section, we investigate decay of the heavy scalars. 
For illustration, we focus on the normal ordering spectrum $M_1<M_2<M_3$ and study decay of the heaviest scalar $S_3$ induced only from the ${\cal O}^4_{ijkl}$ operator Eq.~(\ref{eq:O4-EWSB}) and renormalizable scalar interactions. 
Decay of the second heaviest scalar $S_2$ can be easily translated from that of $S_3$, thanks to the flavor symmetry. 
We cover only the leading order terms in the $\epsilon$ expansion until Section \ref{sec:NNLO}, where effects from higher order terms are discussed. 

In the following subsections, 
we will make various approximations to evaluate lifetimes, and pay a particular attention to scaling of decay widths in terms of model parameters, rather than to accuracy of calculations. 
Hence lifetime calculations given in this section should be regarded as estimates. 
We add that in this MFV framework there are a lot of UV model-dependent ${\cal O}(1)$ coefficients, which absorb calculation uncertainties to some extent. 
Thus our conclusion will not largely be changed by performing precise calculations. 
Incidentally, we have confirmed that our calculation provides an order-of-magnitude estimate, compared with a numerical calculation using MadGraph~\cite{Alwall:2014hca}. 

The dominant decay mode depends on the mass splitting $\Delta M \equiv M_3-M_1$. 
Given that the leading decay vertex Eq.~(\ref{eq:O4-EWSB}) is necessarily accompanied by the top quark due to the flavor symmetry, $S_3$ can decay predominantly into a pair of the top quark and a lighter quark if $\Delta M > m_t$. 
If $\Delta M < m_t$, however, it cannot decay into the on-shell top quark and the dominant decay mode should be four or five-body processes through the off-shell top-quark propagator. 
From Section \ref{sec:S3-S1tu} to \ref{sec:S3-decay_five-body}, 
we elaborate on such multi-body decays produced at the leading order of $\eps$. 
Although these multi-body decays are the leading order in the $\epsilon$ expansion, we will find that three-body processes induced at higher orders of $\eps$ can surpass the leading order ones for $\Delta M \ll m_t$, due in part to strong phase-space suppression in the latter. 
We will assess the higher order processes in Section \ref{sec:NNLO}, then identify parameter spaces for multi-component DM in Section \ref{sec:results}.

\subsection{$S_3 \to S_1 t \bar{u}$}
\label{sec:S3-S1tu}

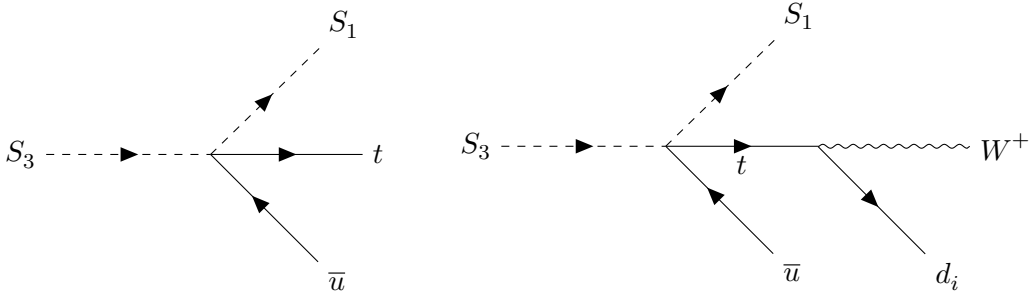
\begin{figure}[t]
    \centering
    \begin{tikzpicture}
        \begin{feynman}
            \vertex (S3) {$S_3$};
            \vertex [right=2.5cm of S3] (intO4);
            \vertex [above right=2cm of intO4] (S1) {$S_1$};
            \vertex [below right=2cm of intO4] (ubar) {$\ol{u}$};
            \vertex [right=2cm of intO4] (t) {$t$};
            
            \diagram* {
            (S3) -- [charged scalar] (intO4) -- [anti fermion] (ubar),
            (intO4) -- [charged scalar] (S1),
            (intO4) -- [fermion] (t),
            };
        \end{feynman}
    \end{tikzpicture}
    \hspace{.5cm}
    \begin{tikzpicture}
        \begin{feynman}
            \vertex (S3) {$S_3$};
            \vertex [right=2.5cm of S3] (intO4);
            \vertex [above right=2cm of intO4] (S1) {$S_1$};
            \vertex [below right=2cm of intO4] (ubar) {$\ol{u}$};
            \vertex [right=2cm of intO4] (intCC);
            \vertex [below right=2cm of intCC] (di) {$d_i$};
            \vertex [right=2cm of intCC] (W) {$W^+$};
            
            \diagram* {
            (S3) -- [charged scalar] (intO4) -- [anti fermion] (ubar),
            (intO4) -- [charged scalar] (S1),
            (intO4) -- [fermion, edge label'=$t$] (intCC) -- [fermion] (di),
            (intCC) -- [photon] (W),
            };
        \end{feynman}
    \end{tikzpicture}        
    \caption{Feynman diagrams for $S_3 \to S_1 t \overline{u}$ (left) and $S_3 \to S_1 d_i \overline{u} W^+$ (right)}
    \label{fig:S3decay1}
\end{figure}

If $\Delta M$ is larger than the top quark mass, 
$S_3$ decays into the on-shell top quark (Fig.~\ref{fig:S3decay1}). 
Since the decay vertex is parameterized by the top Yukawa coupling, this decay mode is naturally dominant if kinematically allowed. 

The squared amplitude for this decay process is given by 
\beq
\sum_{\rm spin,\,color}|\M(S_3\to S_1 t \bar{u})|^2 
	= N_c \(\frac{c_1 m_t}{\Lambda^2}\)^2 (m_{12}^2-m_t^2) \,,
\label{eq:M2-3to1tu}
\eeq
where we ignored the up-quark mass and $N_c=3$ denotes the number of quark colors. 
The invariant masses, $m_{12}^2$ and $m_{23}^2$, are defined 
in terms of the outgoing four-momenta of the decay products,
\beq
m_{12}^2 = (p_1+p_t)^2 \,, \quad
m_{23}^2 = (p_t+p_u)^2 \,.
\eeq
The partial decay width is evaluated by 
\begin{align}
\Gamma(S_3\to S_1 t \bar{u}) 
	& = \frac{N_c}{256\pi^3 M_3^3} \(\frac{c_1 m_t}{\Lambda^2}\)^2
		\int_{m_t^2}^{\(\Delta M\)^2} \d m_{23}^2 
		\frac{(m_{23}^2-m_t^2)^2}{m_{23}^2} \sqrt{\la(M_3^2,M_1^2,m_{23}^2)} \,,
\label{eq:S3-S1tu}
\end{align}
with $\la(\alpha,\beta,\gamma) = \alpha^2+\beta^2+\gamma^2-2(\alpha\beta+\beta\gamma+\alpha\gamma)$. 
While this integral is performed numerically in our analysis, 
it is useful to provide an approximate width for $m_t \ll \Delta M \ll M_3+M_1$. 
In this limit, we have 
\begin{align}
\Gamma(S_3\to S_1 t \bar{u}) 
    & \simeq \frac{N_c\(M_1+M_3\)\(\Delta M\)^5}{960\pi^3 M_3^3} \(\frac{c_1m_t}{\Lambda^2}\)^2 
    =: \Gamma_0 \,.
\label{eq:S3-S1tu-approx}
\end{align}
This is a baseline decay width for $S_3$ and we define it as $\Gamma_0$ for later convenience.

\subsection{$S_3 \to S_1 d_i \bar{u} W$}
\label{sec:S3-decay_four-body}

Below the top threshold $\Delta M \leq m_t$, 
$S_3$ can only decay into the off-shell top quark. 
Then, four-body processes $S_3 \to S_1 d_i \bar{u} W^+$, which are allowed for $m_W+m_{d_i} \leq \Delta M$, take the place of the dominant decay mode (Fig.~\ref{fig:S3decay1}). 
We estimate these decay widths in this subsection. 

For $\Delta M \ll m_t$, we find the squared decay amplitudes, 
\beq
|\bar{\M}|^2 
	:= \sum_{\rm spin, color} |\M(S_3 \to S_1 d_i \bar{u} W^+)|^2
	\simeq N_c \( \frac{y_t \, m_W |V_{ti}|}{\Lambda^2 m_t^2} \)^2 
	\times \frac{8 (p_{d_i} \cdot p_W)^2 (p_u \cdot p_{d_i})}{m_W^2} \,,
\label{eq:M2-buW}
\eeq
where we take $p_u^2 = p_{d_i}^2 = p_W^2 =0$ and 
keep the leading term in the limit where $m_W^2 \ll (p_u \cdot p_{d_i}),\, (p_{d_i} \cdot p_W)$.
The partial decay widths are obtained by integrating the squared amplitudes over four-body phase space. 
Such a integral can be performed numerically or using a public calculation package, such as MadGraph~\cite{Alwall:2014hca}, but we instead estimate the four-body decay widths as 
\begin{align}
\Gamma(S_3 \to S_1 d_i \bar{u} W^+) 
    & \sim \frac{(2\pi)^4}{2M_3} |\bar{\M}|^2 \times \Phi_4(M_3; M_1, 0) \,,
\end{align}
where $\Phi_4(M_3; M_1,0)$ is the four-body phase space 
for only $S_1$ massive and the others massless. 
Using two-cluster decomposition (see Appendix \ref{sec:PhiN} for the detail), 
we have $\Phi_4 (M_3; M_1, 0)$ in the form, 
\begin{align}
\Phi_4 (M_3; M_1, 0) 
	& = \frac{M_3^4}{393216\pi^9} \, f_3(M_1^2/M_3^2) \,,
\end{align}
where 
\begin{align}
f_3 (v) & \simeq \frac{1}{10} (1-v)^5  \quad \text{for $v\simeq1$} \,.
\end{align}
For $\Delta M \ll M_1+M_3$, we find 
\begin{align}
\Gamma(S_3 \to S_1 d_i \bar{u} W^+) 
	& \sim \frac{N_c\(\Delta M\)^{11}}{414720 \pi^5 M_3^2} 
	\( \frac{y_t \, |V_{ti}|}{\Lambda^2 m_t^2} \)^2 \,.
\end{align}
Here, we replaced the scalar products of the final-state momenta in $|\bar{\M}|^2$ with their mean values. 
Concretely, using the energy-momentum conservation, 
\beq
\(\Delta M\)^2
	\simeq (p_3-p_1)^2 
	= (p_{d_i}+p_W+p_u)^2 
	\simeq 2 (p_{d_i}\cdot p_W + p_{d_i} \cdot p_u + p_u \cdot p_W) \,,
\eeq
and symmetry among $p_{d_i,u,W}$, 
we approximate\footnote{If we evaluate the decay width for $S_3 \to S_1 t \bar{u}$ in a similar way, the width is overestimated by a factor of 2.5.}
\beq
p_{d_i} \cdot p_W 
\sim p_{d_i} \cdot p_u 
\sim p_u \cdot p_W 
\sim \frac{\(\Delta M\)^2}{6} \,.
\eeq
The ratios of these four-body decay widths (or equally the branching ratios) are determined only by the CKM matrix elements:
\begin{align}
\Gamma(S_3 \to S_1 b \bar{u} W^+)  
    &\simeq \left|\frac{V_{tb}}{V_{ts}}\right|^2 \Gamma(S_3 \to S_1 s \bar{u} W^+) 
    \simeq \left|\frac{V_{tb}}{V_{td}}\right|^2 \Gamma(S_3 \to S_1 d \bar{u} W^+) \,.
\end{align}
This relation is robust and independent of whether we evaluate the phase-space integral numerically or make just an estimate like above.

\subsection{$S_3 \to S_1 d_i \bar{u} f \bar{f'}$}
\label{sec:S3-decay_five-body}

\begin{figure}[t]
    \centering
    \begin{tikzpicture}
        \begin{feynman}
            \vertex (S3) {$S_3$};
            \vertex [right=2.5cm of S3] (intO4);
            \vertex [above right=2cm of intO4] (S1) {$S_1$};
            \vertex [below right=2cm of intO4] (ubar) {$\ol{u}$};
            \vertex [right=2cm of intO4] (intCC1);
            \vertex [below right=2cm of intCC1] (di) {$d_i$};
            \vertex [right=2cm of intCC1] (intCC2);
            \vertex [right=2cm of intCC2] (f) {$f$};
            \vertex [below right=2cm of intCC2] (f') {$\ol{f'}$};
            
            \diagram* {
            (S3) -- [charged scalar] (intO4) -- [anti fermion] (ubar),
            (intO4) -- [charged scalar] (S1),
            (intO4) -- [fermion, edge label'=$t$] (intCC1) -- [fermion] (di),
            (intCC1) -- [photon, edge label=$W$] (intCC2) -- [fermion] (f),
            (intCC2) -- [anti fermion] (f'),
            };
        \end{feynman}
    \end{tikzpicture}        
    \caption{Feynman diagram for five-body decay processes $S_3 \to S_1 d_i \bar{u} f \ol{f'}$.}
    \label{fig:S3decay2}
\end{figure}
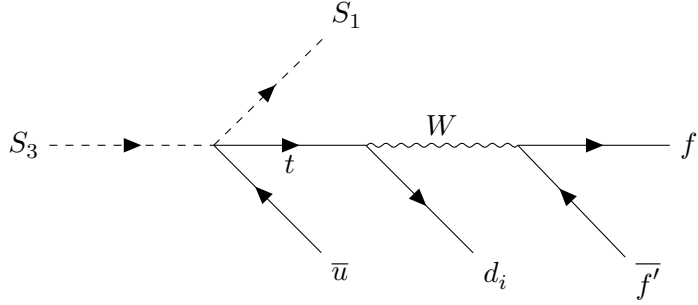

As the mass splitting $\Delta M$ gets smaller than the $W$ boson mass, 
even $S_3 \to S_1 d_i \bar{u} W$ decays are kinematically forbidden. 
In this case, five-body processes via the off-shell $W$ exchange (Fig.~\ref{fig:S3decay2}) start to dominate the $S_3$ decay. 
Here, we estimate these decay widths assuming the mass splitting is larger than 1\,GeV so that we can evaluate the widths by parton-level calculation. 

For $\Delta M \ll m_W$, the squared decay amplitudes are given by 
\begin{align}
|\bar{\M}|^2 
	& = \sum_{\rm spin, color} |\M(S_3 \to S_1 d_i \bar{u} f \bar{f'})|^2 \nonumber\\
	& = \( \frac{2|V_{ti}| |U_{ff'}|}{\Lambda^2 m_t v^2} \)^2 
		\times 32 N_c N_{c,f} (p_b \cdot p_f) \times \nonumber\\
	& \quad \left\{ (p_{d_i} \cdot p_{f'} + p_f \cdot p_{f'}) (p_{d_i} \cdot p_u + p_u \cdot p_f) 
		- (p_{d_i} \cdot p_f) (p_u \cdot p_{f'}) \right\} \,,
\label{eq:M2-buff'}
\end{align}
where $U_{ff'} = V_{ff'}$ for quarks and $U_{ff'} = \delta_{ff'}$ for leptons, and $N_{c,f}$ is the number of colors for a fermion $f$.
The partial decay widths are estimated in a similar way to the four-body case by making an approximation, 
\beq
\Gamma(S_3 \to S_1 d_i \bar{u} f \bar{f'}) 
	\sim \frac{(2\pi)^4}{2M_3} |\bar{\M}|^2 \times \Phi_5(M_3; M_1, 0) \,,
\eeq
where $\Phi_5(M_3; M_1,0)$ is the five-body phase space 
for only $S_1$ massive and the others massless and 
given explicitly by
\begin{align}
\Phi_5 (M_3; M_1, 0) 
	& = \frac{M_3^6}{75497472\pi^{11}} \, f_4(M_1^2/M_3^2) \,,
\end{align}
with 
\begin{align}
f_4 (v) & \simeq \frac{1}{35} (1-v)^7  \quad \text{for $v\simeq1$} \,.
\end{align}
Then, we find the decay widths for $\Delta M \ll M_1+M_3$, 
\begin{align}
\Gamma(S_3 \to S_1 d_i \bar{u} f \bar{f'}) 
	& \sim \frac{N_c N_{c,f} \(\Delta M\)^{13}}{11612160 \pi^7 M_3^2} 
		\( \frac{|V_{ti}| |U_{ff'}|}{\Lambda^2 m_t v^2} \)^2 \,.
\end{align}
Here, we applied a similar approximation to the four-momentum products for the light final-state particles, i.e.
\beq
p_A \cdot p_B \sim \frac{\(\Delta M\)^2}{12} \,,
\eeq
for $A,B=d_i,u,f,f'$. 
In addition to a small numerical factor $1/(11612160\pi^7)$, 
the decay widths are proportional to the power of huge scale differences originating from $\Delta M \ll v, M_3, \Lambda$. 
These suppression factors rapidly reduce the widths as $\Delta M$ becomes small. 
It is easy to find again a close relation among these five-body decay widths, 
\begin{align}
\Gamma(S_3 \to S_1 b \bar{u} f \bar{f'})
    & \simeq \left|\frac{V_{tb}}{V_{ts}}\right|^2 \Gamma(S_3 \to S_1 s \bar{u} f \bar{f'}) 
    \simeq \left|\frac{V_{tb}}{V_{td}}\right|^2 \Gamma(S_3 \to S_1 d \bar{u} f \bar{f'}) \,,
\end{align}
if these decay processes are kinematically allowed.

\subsection{Higher-order contributions in the MFV expansion}
\label{sec:NNLO}

We have studied the leading order effects in the MFV expansion up to here, and found that those contribution to the heavy scalar decay receives strong phase-space suppression for a small mass splitting. 
That encourages us to evaluate higher order contribution that has extra suppression from $\eps$ but evades strong phase-space suppression. 
In this section, we study the impact of higher order terms on the heavy scalar decay.

For the scalar mass and Higgs portal coupling, 
higher order corrections are taken into account by 
\begin{align}
m_S^2 \, \1 
	& \to m_S^2 \[ a_0 \, \1 + \eps \, a_1 \(Y_u^\dag Y_u\)
		+ \eps^2 \( a_2  Y_u^\dag Y_d Y_d^\dag Y_u + a_2' (Y_u^\dag Y_u)^2 \) 
		+ {\cal O}(\eps^3) \] \,, \label{eq:MS_NLO}\\
\la \, \1 
	& \to \la \[ b_0 \, \1 + \eps \, b_1 \(Y_u^\dag Y_u\) 
		+ \eps^2 \( b_2 Y_u^\dag Y_d Y_d^\dag Y_u + b_2' (Y_u^\dag Y_u)^2 \) 
		+ {\cal O}(\eps^3) \] \,, \label{eq:lam_NLO}
\end{align}
where $a$'s and $b$'s are ${\cal O}(1)$ coefficients. 
The first flavor off-diagonal elements arise from the $\eps^2$ terms. 
Given that 
$(Y_u)_{ij} = (V^\dag \hat{Y}_u)_{ij}$ and 
$(Y_d)_{ij} = (\hat{Y}_d)_{ij} = y_d^i \, \delta_{ij}$, 
the scalar mass terms are expressed by 
\beq
{\cal L}_{\rm mass} = - S_i^* (M_S^2)_{ij} S_j \,,
\quad 
M_S^2 = 
	\( \begin{array}{ccc} 
		M_1^2  & \Delta M_{12}^2 & \Delta M_{13}^2  \\
		\Delta M_{21}^2 & M_2^2  & \Delta M_{23}^2  \\
		\Delta M_{31}^2  & \Delta M_{32}^2 & M_3^2 \\
	\end{array} \) \,,
\eeq
where $\Delta M_{ij}^2 = \eps^2 (a_2 m_S^2+ b_2 \la v^2/2) \, y_u^i V_{ik} (y_d^k)^2  V_{jk}^* y_u^j$. 
The mass matrix is diagonalized by a unitary matrix $U_S$, 
which is given for $\eps \ll 1$ by 
\beq
S_i \to (U_S)_{ij} S_j \,,\quad\quad
U_S \simeq \( 
        \begin{array}{ccc} 
		1 & - \theta_{12} & - \theta_{13} \\
		\theta_{12} & 1 & - \theta_{23} \\
            \theta_{13} & \theta_{23} & 1 \\
	\end{array} \) \,,
\eeq
where we approximate $|\theta_{ij}| \ll 1$. 
The mixing angle between $S_i$ and $S_j$ is given by 
\beq
\theta_{ij} 
    \simeq \frac{\Delta M_{ij}^2}{M_i^2-M_j^2} 
    = \eps \, R\, \frac{y_u^i V_{ik} (y_d^k)^2  V_{jk}^* y_u^j}{(y_u^i)^2-(y_u^j)^2} \,,
\eeq
with $M_i^2-M_j^2 = \eps \, [(y_u^i)^2-(y_u^j)^2] \, (a_1 m_S^2 + b_1 \la v^2/2)$ and 
\beq
R \equiv \frac{a_2 m_S^2 + b_2 \la v^2/2}{a_1 m_S^2 + b_1 \la v^2/2} = {\cal O}(1) \,.
\eeq
Note that although the off-diagonal elements in the mass matrix appear at the order of $\eps^2$, the scalar mixing angle is of the order of $\eps$ because the mass splitting is generated at the order of $\eps$.

\subsubsection{Scalar mixing}
\label{sec:decay_via_mixing}

The scalar mixing induces new decay modes that do not appear at the order of $\eps^0$. 
At the order of $\eps$, only the $c_1$ term generates such new decay modes via the scalar mixing.\footnote{Flavor off-diagonal interactions like $\bar{u} u S_1^* S_3$ do not stem from the $c_2$ term at the order of $\eps$, even if the scalar mixing is present. 
This is understood from the fact that $(S_i^* S_i)$ is invariant under the mass diagonalization $S_i \to (U_S)_{ij} S_j$}
In the mass basis, the pertinent interaction Lagrangian is given by 
\begin{align}
\L_{d=6} 
	& = \frac{c_1}{\Lambda^2} \[ \bar{u}_k (m_k P_R + m_l P_L) u_l  \(S_i^* (U_S^*)_{li} (U_S)_{kj} S_j \) \] \,.
\label{eq:O4-mixing}
\end{align}
We are particularly interested in $S_3$ decay into $u, c$ quarks. 
Taking $j=3$ and $k,l\neq3$, 
the largest contribution comes from $i=l$, yielding 
\begin{align}
\Gamma(S_3\to S_i u_k \bar{u_i}) 
    & \simeq \Gamma_0 \times \left|\theta_{i3}\right|^2 \frac{(m_u^i)^2 + (m_u^k)^2}{m_t^2} \,.
\label{eq:S3toS1uu-mixing}
\end{align}
Here, we used for a small mixing angle (or equally a small $\eps$) 
\beq
(U_S)_{ij} 
    \simeq 
        \left\{ 
        \begin{array}{ll} 
            1 & (i=j) \\ 
            - \theta_{ij} \simeq - (U_S^*)_{ji} & (i\neq j)
        \end{array}
        \right. \,.
\label{eq:US_approx}
\eeq
Since $\theta_{i3} \propto \eps$, the decay width is suppressed by $\eps^2$ but without extra phase-space suppression, compared with the reference three-body width $\Gamma_0$, Eq.~(\ref{eq:S3-S1tu-approx}). 
Thus, this process might be as large as the four- or five-body decays discussed in Sections \ref{sec:S3-decay_four-body} and \ref{sec:S3-decay_five-body}, depending on the values of $\eps$ and $\Delta M$. 

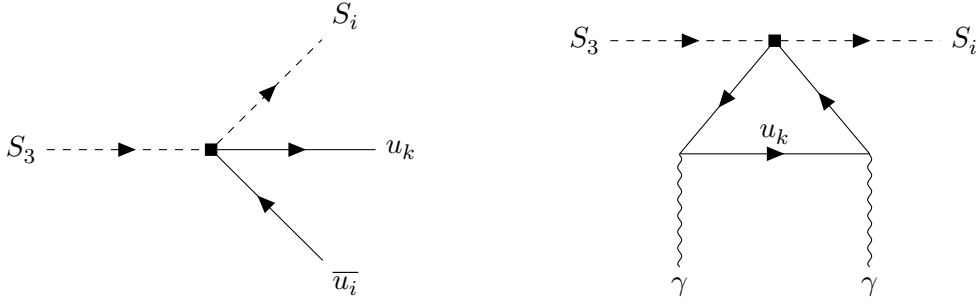
\begin{figure}[t]
    \centering
    \begin{tikzpicture}
        \begin{feynman}
            \vertex [square dot] (intO4) {};
            \vertex [left=2.5cm of intO4](S3) {$S_3$};
            \vertex [above right=2.5cm of intO4] (Si) {$S_i$};
            \vertex [right=2.5cm of intO4] (uk) {$u_k$};
            \vertex [below right=2.5cm of intO4] (uibar) {$\bar{u_i}$};
            
            \diagram* {
            (S3) -- [charged scalar] (intO4) -- [charged scalar] (Si),
            (intO4) -- [fermion] (uk),
            (intO4) -- [anti fermion] (uibar),
            };
        \end{feynman}
    \end{tikzpicture}
    \hspace{1.5cm}
    \begin{tikzpicture}
        \begin{feynman}
            \vertex [square dot] (intO4) {};
            \vertex [left=2.5cm of intO4](S3) {$S_3$};
            \vertex [right=2.5cm of intO4] (Si) {$S_i$};
            \vertex [below=1.5cm of intO4] (pv);
            \vertex [left=1.25cm of pv] (intEM1);
            \vertex [right=1.25cm of pv] (intEM2);
            \vertex [below=1.5cm of intEM1] (photon1) {$\gamma$};
            \vertex [below=1.5cm of intEM2] (photon2) {$\gamma$};
            
            \diagram* {
            (S3) -- [charged scalar] (intO4) -- [charged scalar] (Si),
            (intO4) -- [fermion] (intEM1) --  
            [fermion, edge label=$u_k$] (intEM2) -- [fermion] (intO4),
            (intEM1) -- [photon] (photon1), 
            (intEM2) -- [photon] (photon2), 
            };
        \end{feynman}
    \end{tikzpicture}
    \caption{
    (Left) Feynman diagram for $S_3 \to S_i u_k \bar{u_i}$ decay, which arises from the scalar mixing at the order of $\eps$. The black square dot means the ${\cal O}^4_{ijkl}$ vertex at the order of $\eps$. 
    (Right) Feynman diagram for $S_3 \to S_i \gamma \gamma$ decay through the same interaction vertex at the order of $\eps$. 
    }
    \label{fig:S3decay-loop}
\end{figure}

By closing the quark lines for the $k=l$ interaction vertices in Eq.~(\ref{eq:O4-mixing}), 
we have $S_j \to S_i \gamma\gamma$ decay at one-loop level (Fig.~\ref{fig:S3decay-loop}). 
The decay amplitude is given by 
\beq
i{\cal M}(S_j \to S_i \gamma\gamma) 
    = i \frac{c_1}{\Lambda^2} \({\cal A}_{\gamma\gamma}\)_{ij} \,
        \eps^*_\mu(p) \eps^*_\nu(q) 
        \[(p\cdot q) g^{\mu\nu} - p^\nu q^\mu\] \,,
\eeq
where $p$ and $q$ denote outgoing four-momenta of two final-state photons and 
\begin{align}
\({\cal A}_{\gamma\gamma}\)_{ij} 
    & = \frac{N_c \alpha Q_u^2}{2\pi} \sum_{k=1,2,3} (U_S^*)_{ki} (U_S)_{kj} F_{1/2}(\tau_k) \,,
\end{align}
with $\tau_k=4(m_u^k)^2/(2p\cdot q)$. 
The loop function is well-known in the context of the Higgs diphoton decay (see e.g. \cite{Gunion:1989we}), and given by 
\beq
F_{1/2}(\tau) = - 2 \tau \[1+\(1-\tau\)f(\tau)\] \,,
\label{eq:F1/2}
\eeq
where 
\beq
f(\tau) 
    = \left\{
    \begin{array}{ll}
        \arcsin^2(\sqrt{1/\tau})    & (\tau \geq 1) \\
        -\frac{1}{4}\[\ln(\eta_{+}/\eta_{-}) -i\pi\]^2 & (\tau < 1)
    \end{array}
    \right.\,,
    \label{eq:ftau}
\eeq
with $\eta_\pm = 1\pm\sqrt{1-\tau}$. 
It is useful to show two limits of $F_{1/2}$: 
$F_{1/2}(\tau\to\infty)=-4/3$ and $F_{1/2}(\tau\to0)=0$. 
It follows from these two limits that 
the top loop contribution dominates $S_3 \to S_1 \gamma \gamma$ decay for $m_u \ll \Delta M \ll m_t$. 
The charm loop contribution to $({\cal A}_{\gamma\gamma})_{13}$ is proportional to $\theta_{12}^*\theta_{23} \propto \eps^2$ and thus negligible. 
For $m_u \ll \Delta M \ll m_t$, therefore, the decay width approximates to 
\begin{align}
\Gamma(S_3 \to S_1 \gamma \gamma)_{\rm loop}
    & \simeq \frac{M_1+M_3}{256\pi^3 M_3^3} \left|\theta_{13}\right|^2 
        \(\frac{c_1 N_c \alpha Q_u^2}{3\pi\Lambda^2}\)^2 \times 
        \frac{16}{105} \(\Delta M\)^7 \,.
\end{align}
Note that because of unitarity of the scalar mixing matrix, the up- and top-quark contributions cancelled each other $({\cal A}_{\gamma\gamma})_{13} \simeq 0$ for $\Delta M \ll m_u$, where the up-quark contribution is also saturated with its asymptotic value $F_{1/2}(\tau_u)\simeq-4/3$. 
However, the up-quark contribution is highly sensitive to how we treat the up-quark mass, resulting in a large calculation uncertainty. 
If we use current quark mass, $({\cal A}_{\gamma\gamma})_{13} \simeq 0$ below ${\cal O}(\MeV)$, while if we take it as constituent quark mass, $({\cal A}_{\gamma\gamma})_{13} \simeq 0$ below ${\cal O}(100\,\MeV)$.

As the mass splitting becomes below $\sim 1\,\GeV$, we have to consider decay into hadrons rather than partons, Eq.~(\ref{eq:S3toS1uu-mixing}). 
To evaluate such hadronic decays, we first derive effective Lagrangian at the QCD scale by taking $k=l$ in Eq.~(\ref{eq:O4-mixing}) and integrating out charm and top quarks. 
The relevant effective interactions are given by 
\begin{align}
\L_{\rm eff,\,mixing} 
	& = \frac{c_1}{\Lambda^2} 
			\[ m_u (U_S^*)_{1i} (U_S)_{1j} \, \(\bar{u} u\) 
				- \sum_{k=2,3} (U_S^*)_{ki} (U_S)_{kj} \frac{\alpha_s}{12\pi} G_{\mu\nu}^a G^{\mu\nu a} 
			\] \(S_i^* S_j\) \,,
\end{align}
where the second term arises from charm and top loops. 
Using unitarity of the scalar mixing matrix, we can rewrite the second term in the square bracket with 
\beq
\sum_{k=2,3} (U_S^*)_{ki} (U_S)_{kj} \frac{\alpha_s}{12\pi} G_{\mu\nu}^a G^{\mu\nu a} 
	= \[ \delta_{ij} - (U_S^*)_{1i} (U_S)_{1j} \] \frac{\alpha_s}{12\pi} G_{\mu\nu}^a G^{\mu\nu a} \,,
\eeq
which yields for $i \neq j$ 
\begin{align}
\L_{\rm eff,\,mixing} 
	& = \frac{c_1}{\Lambda^2} 
			\( m_u \, \bar{u} u + \frac{\alpha_s}{12\pi} G_{\mu\nu}^a G^{\mu\nu a} \) (U_S^*)_{1i} (U_S)_{1j}  \(S_i^* S_j\) \,.
\end{align}
In the leading order chiral perturbation, we find hadronic matrix elements for quarks and gluon \cite{Vainshtein:1980ea, Voloshin:1980zf, Novikov:1980fa, Voloshin:1985tc, Grinstein:1988yu}, 
\begin{align}
\langle \pi^a(p) \pi^b(q) | m_u \bar{u} u | 0 \rangle 
	& = \frac{1}{2} \delta^{ab} m_\pi^2 \,,\\
\langle \pi^a(p) \pi^b(q) | \frac{9\alpha_s}{8\pi} G_{\mu\nu}^a G^{\mu\nu a} | 0 \rangle 
	& = - \delta^{ab} (s + m_\pi^2) \,,
\end{align}
where we use the isospin symmetry. 
For $m_\pi \ll \Delta M$, the decay widths are approximately evaluated as 
\begin{align}
\Gamma(S_j \to S_i \pi^a \pi^b) 
	& \simeq \frac{\delta^{ab}(M_j+M_i)}{512\pi^3 M_j^3} \(\frac{2c_1}{27\Lambda^2}\)^2 
		|(U_S^*)_{1i} (U_S)_{1j}|^2 \times \frac{16}{105} (M_j-M_i)^7 \,.
\end{align}
Using Eq.~(\ref{eq:US_approx}), we notice $\Gamma(S_3 \to S_2 \pi^a \pi^b) \propto |\theta_{12}^*\theta_{13}|^2 \propto \eps^4$ and it has a minor effect in our order-counting. 
Note that there should be large calculation uncertainties due to significant final-state interactions of pions \cite{Raby:1988qf, Grinstein:1988yu, Truong:1989my, Chivukula:1989ds}. 
For $2m_\pi < \Delta M <1\,\GeV$, the decay widths could receive an enhancement by as much as a factor of 10. 
Nonetheless, these calculation uncertainties would not change our basic conclusion.

\subsubsection{Higgs portal contribution}

Higher order terms of the Higgs portal coupling also induces other new decay modes. 
In the mass basis of $S_i$, 
flavor off-diagonal interactions to the Higgs boson stem from the $\eps^2$ terms, 
\begin{align}
V(H,S) 
    & \supset \lambda v \left\{ \eps \, b_1 \sum_k (y_u^k)^2 \[(\theta^*)_{ki} \delta_{kj} + \delta_{ki} (\theta)_{kj}\] + \eps^2 \, b_2 \, y_u^i y_u^j \sum_ k (y_d^k)^2 V_{ik} V^*_{jk} \right\} h S_i^* S_j \,,\nonumber\\
    & \simeq \eps^2 \lambda v A \(y_u^i y_u^j \sum_ k (y_d^k)^2 V_{ik} V^*_{jk} \) h S_i^* S_j
    =: \eps^2 \lambda_{ij} v h S_i^* S_j \,,
\label{eq:lambda_ij}
\end{align}
with
\beq
\lambda_{ij} \equiv \lambda A \(y_u^i y_u^j \sum_ k (y_d^k)^2 V_{ik} V^*_{jk} \) \,,
\eeq
where $A$ denotes a model-dependent ${\cal O}(1)$ coefficient defined by 
\beq
A \equiv - b_1 \frac{a_2 m_S^2 + b_2 \la v^2/2}{a_1 m_S^2 + b_1 \la v^2/2} + b_2 \,.
\eeq
The first term in $A$ comes from a combination of the scalar mixing and the flavor diagonal couplings, both at the order of $\eps$. 
Meanwhile, the second term stems purely from the flavor off-diagonal elements of the Higgs portal coupling. 
These two terms equally contribute to new decay modes. 

First, Eq.~(\ref{eq:lambda_ij}) induces two-body decay $S_3 \to S_1 h$, whose width is given by 
\beq
\Gamma(S_3 \to S_1 h) 
    \simeq \frac{\eps^4 \left|\lambda_{13}\right|^2 v^2\beta}{16\pi M_3} \,,
\eeq
with 
\beq
\beta = \sqrt{1-\frac{(M_1+m_h)^2}{M_3^2}} \sqrt{1-\frac{(M_1-m_h)^2}{M_3^2}} \,.
\eeq
This decay is possible only for $\Delta M>m_h$. 
Compared with $S_3 \to S_1 t \bar{u}$, the decay width for $S_3 \to S_1 h$ is suppressed by $\eps^4$ and seems not to be dominant. 
This is, however, two-body decay and has no suppression from the UV cutoff scale $\Lambda$, 
so it can have some impact for a large $\Lambda$.

For $\Delta M < m_h$, $S_3 \to S_1 h$ is forbidden and off-shell Higgs-mediated processes begin to dominate. 
For $\Delta M \lesssim {\cal O}(v)$, effective interactions to the SM fields are given by 
\beq
\L_{\rm eff,\,higgs}
	 = \frac{\eps^2 \lambda_{13}}{m_h^2} \(S_1^* S_3\) \(\sum_{f} m_f \bar{f} f 
		+ \frac{\alpha_s F_g}{16\pi} G_{\mu\nu}^a G^{\mu\nu a} 
		+ \frac{\alpha F_\gamma }{8\pi} F_{\mu\nu} F^{\mu\nu} \) \,,
\eeq
where $f$ runs over all charged leptons and quarks that are lighter than $\Delta M/2$.
The effective couplings to gluon and photon fields are given by 
\begin{align}
F_g & = \sum_q F_{1/2}(\tau_q) \, \Theta(2m_q-\Delta M) \,,\\
F_\gamma & = F_1(\tau_W) + \sum_f N_{c,f} q_f^2 \, F_{1/2}(\tau_f) \, \Theta(2m_f-\Delta M)\,,
\end{align}
with $N_{c,f}$ and $q_f$ being the color and electric charge for a fermion $f$ and $\tau_i=4m_i^2/(2p\cdot q)$ for a particle $i$ in the loop. 
The loop functions are given by Eq.~(\ref{eq:F1/2}) and 
\begin{align}
F_1(\tau) & = 2 + 3\tau + 2\tau(2-\tau) f(\tau) \,   \qquad \mbox{with $F_1(\tau\to\infty)=7$} \,,
\end{align}
where $f(\tau)$ is in Eq.~(\ref{eq:ftau}).
$S_3$ can decay into $S_1$ plus a pair of fermions, gluons or photons. 
We approximately obtain the partial widths for those decay processes, 
\begin{align}
\Gamma(S_3 \to S_1 f \bar{f})
    & \simeq \frac{N_{c,f}\(M_1+M_3\)\(\Delta M\)^5}{480\pi^3 M_3^3} 
            \(\frac{\eps^2|\lambda_{13}|m_f}{m_h^2}\)^2 \,,\\
\Gamma(S_3 \to S_1 g g)
    & \simeq \frac{M_1+M_3}{512\pi^3 M_3^3}
            \(\frac{\eps^2|\lambda_{13}|}{m_h^2}\)^2
            \(\frac{\alpha_s}{2\pi} \frac{4}{3}N_h\)^2 \times 
            \frac{15}{106} \(\Delta M\)^7 \,,\\
\Gamma(S_3 \to S_1 \gamma \gamma) 
	& \simeq \frac{M_1+M_3}{1024\pi^3 M_3^3}
            \(\frac{\eps^2|\lambda_{13}|}{m_h^2}\)^2 
            \(\frac{\alpha}{2\pi} f_\gamma\)^2 \times 
            \frac{15}{106} \(\Delta M\)^7 \,,
\end{align}
where $N_h$ denotes the number of quarks heavier than $\Delta M/2$ and 
\beq
f_\gamma = 7-\frac{4}{3} \sum_f N_{c,f} q_f^2 \Theta(2m_f-\Delta M) \,.
\eeq
To obtain these approximate widths, we assumed $m_f \ll \Delta M \ll M_3+M_1$ and used the corresponding values of 
$F_1(\tau\to\infty)$ and $F_{1/2}(\tau\to\infty)$.
We also ignored the interference with the $c_1$-induced terms, obtained in Section \ref{sec:decay_via_mixing}. 
The cutoff scale $\Lambda$ is replaced with the Higgs boson mass in these processes.  
This means that $\epsilon$ or $\lambda$ must be small enough to make $S_3$ long-lived. 
We will see how small $\eps$ and $\lambda$ should be in the next section. 

Below 1\,GeV, we have to consider hadronic decay. 
The effective Lagrangian at 1\,GeV consists of light quarks and gluon, 
\beq
\L_{\rm eff,\,higgs} 
	= \frac{\eps^2 \lambda_{ij}}{m_h^2} \(S_i^* S_j\) 
		\( \sum_{q=u,d,s} m_q \bar{q} q - N_h \frac{\alpha_s}{12\pi} G_{\mu\nu}^a G^{\mu\nu a} \) \,,
\eeq
where $N_h=3$ counts $c,b,t$ quarks. 
Using the matrix elements for the light quarks and gluon, evaluated in the leading order chiral perturbation, 
\beq
\langle \pi^a(p) \pi^b(q) | m_u \bar{u} u + m_d \bar{d} d | 0 \rangle = \delta^{ab} m_\pi^2 \,,\quad
\langle \pi^a(p) \pi^b(q) | m_s \bar{s} s | 0 \rangle = 0 \,,
\eeq
we obtain an approximate form of the partial width for $S_3 \to S_1 \pi^a \pi^b$, 
\beq
\Gamma(S_3 \to S_1 \pi^a \pi^b) 
    \simeq \frac{\delta^{ab}\(M_1+M_3\)}{512\pi^3 M_3^3} 
            \(\frac{2\eps^2|\lambda_{13}|}{9m_h^2}\)^2 \times 
            \frac{15}{106} \(\Delta M\)^7 \,.
\eeq

\begin{figure}[t]
    \centering
    \begin{tikzpicture}
        \begin{feynman}
            \vertex (S3) {$S_3$};
            \vertex [crossed dot, right=2.5cm of S3] (i1) {};
            \vertex [above right=2cm of i1] (S1) {$S_1$};
            \vertex [below right=2cm of i1] (h) {$h$};
            
            \diagram* {
            (S3) -- [charged scalar] (i1) -- [charged scalar] (S1),
            (i1) -- [scalar] (h),
            };
        \end{feynman}
    \end{tikzpicture}
    \hspace{.5cm}
    \begin{tikzpicture}
        \begin{feynman}
            \vertex (S3) {$S_3$};
            \vertex [crossed dot, right=2.5cm of S3] (i1) {};
            \vertex [above right=2cm of i1] (S1) {$S_1$};
            \vertex [below right=2cm of i1] (h);
            \vertex [above right=2cm of h] (f) {$f$};
            \vertex [below right=2cm of h] (fbar) {$\bar{f}$};
            
            \diagram* {
            (S3) -- [charged scalar] (i1) -- [charged scalar] (S1),
            (i1) -- [scalar, edge label'=$h$] (h) -- [fermion] (f),
            (h) -- [anti fermion] (fbar),
            };
        \end{feynman}
    \end{tikzpicture}
    \caption{Feynman diagrams for $S_3 \to S_1 h$ (left) and $S_3 \to S_1 h^* \to S_1 f \bar{f}$ (right). The crossed dot means the Higgs portal vertex appearing at the order of $\eps^2$.}
    \label{fig:S3decay_higgs}
\end{figure}
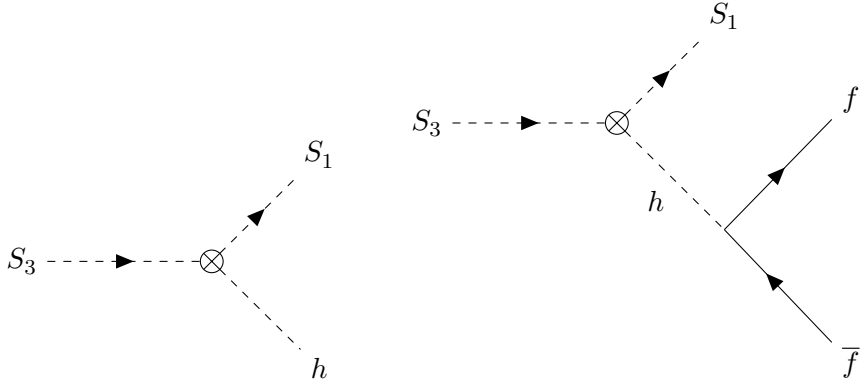

\subsubsection{Dimension-6 operators}

For the ${\cal O}^4_{ijkl}$ operator, higher order corrections correspond to taking 
\begin{align}
c^4_{ijkl} 
    & \sim \eps^2 \[ 
        c\,(Y_u Y_u^\dag Y_d Y_d^\dag Y_u)_{il} \, \delta_{kj} 
        + c'\,(Y_u)_{ij} (Y_u^\dag Y_d Y_d^\dag Y_u)_{kl} 
        + c''\,(Y_u)_{il} (Y_u^\dag Y_d Y_d^\dag Y_u)_{kj}\] \,,
\label{eq:c4-NLO}
\end{align}
where we suppress irrelevant terms that appear at the same order but do not lead to new decay modes. 
There also exists flavor off-diagonal contribution through a combination of the scalar mass mixing and $c^4_{ijkl} \sim {\cal O}(\eps)$ terms, but we ignore it here since it has the same coupling scaling. 
From Eq.~(\ref{eq:c4-NLO}), we have 
\begin{align}
\L_{d=6} 
	& \sim \frac{\eps^2}{\Lambda^2} 
            \sum_{i,j,k,l} \(y_u^i y_u^j (y_d^k)^2 V_{ik} V^*_{jk}\) 
            \left\{ c \, m_u^i \, (\bar{u}_{Li} \, u_{Rl}) (S_l^* S_j) \right. \nonumber\\
        & \quad \left.{}
            + c' m_u^l \, (\bar{u}_{Ll} \, u_{Rl}) (S_i^* S_j) + c'' m_u^l \, (\bar{u}_{Ll} \, u_{Rj}) (S_i^* S_l)\right\} + \hc \,,
\label{eq:o4-NLO}
\end{align}
where all scalar and quark fields are in the mass basis. 
These interactions enable new three-body decay modes that exclude the top quark in the final states. 
The partial widths for such decay processes is evaluated using Eqs.~(\ref{eq:S3-S1tu-approx}) and (\ref{eq:o4-NLO}) as 
\begin{align}
\Gamma(S_3\to S_1 u_i \bar{u}_j) 
    & \simeq \Gamma_0 \times \eps^4
        \left| c \, \delta_{1j} (y_u^i)^2 \sum_k (y_d^k)^2 V_{ik} V^*_{tk} 
        + c' \delta_{ij} y_u^i y_u \sum_{k} (y_d^k)^2 V_{uk} V_{tk}^*\right|^2 \,,
\label{eq:S3-S1uu-dim6}
\end{align}
where $i,j\neq 3$. 
This is suppressed by $\eps^4$ and the light quark Yukawa couplings as well as the CKM matrix elements, compared with the baseline $S_3 \to S_1 t \bar{u}$ decay width $\Gamma_0$. 
Therefore, these are expected not to surpass the other processes already discussed above, unless there is a significant accidental cancellation among the ${\cal O}(1)$ coefficients in the other decay modes.

\section{Parameter spaces for multi-component dark matter}
\label{sec:results}

We are ready to survey model parameter spaces and identify where the heavy components are also DM. 
Our benchmark model is characterized mostly by four parameters, 
\beq
M_1, \quad \Lambda, \quad \lambda, \quad \eps 
\eeq
which control not only the heavy scalar decays but DM physics. 
Magnitude of these four parameters is arbitrary as long as they respect perturbative unitarity ($\lambda \leq 4\pi$) and validity of the EFT ($M_1 \leq \Lambda$). 
As for $\eps$, one can take any value in principle, but we assume $\eps \ll 1$ in order to justify the MFV expansion. 
Note that even if taking $\eps=0$ at the beginning, loop diagrams with the weak interactions necessarily generate higher order terms of $Y_u$ and $Y_d$. 
This suggests that there is a minimum (or natural) value for $\eps$, which is obtained by identifying $\eps$ as a loop factor: $\eps\sim1/(4\pi)^2\sim10^{-2}$\,--\,$10^{-3}$. 
We therefore take $\eps=10^{-2}$ as our benchmark value. 

The mass differences among $S_i$ are generated by $\eps$. 
We compute two mass differences, 
$\Delta M=M_3-M_1$ and $\delta M=M_2-M_1$, 
by numerically solving Eq.~(\ref{eq:DelM}). 
For a small $\eps$, these are approximately given by 
\beq
\eps \simeq \frac{2\Delta M}{y_t^2 M_1} 
    \simeq \frac{2\delta M}{y_c^2M_1} \,,
\eeq
which means that the mass splitting between $S_3$ and $S_1$ is around 0.5\,\% for $\eps=10^{-2}$. 
We ignore flavor-diagonal corrections to the Higgs portal coupling $\lambda$, and take $\lambda_{hSi}=\lambda$ in the following analysis. 
This choice does not influence our results, since the leading Higgs-mediated decays are already suppressed by $\eps^4$ and further corrections are insignificant. 
We also set all UV-model dependent ${\cal O}(1)$ coefficients to unity: $c_1=c_2=A=R=1$.

\begin{figure}[t]
\centering
    \includegraphics[width=0.48\textwidth]{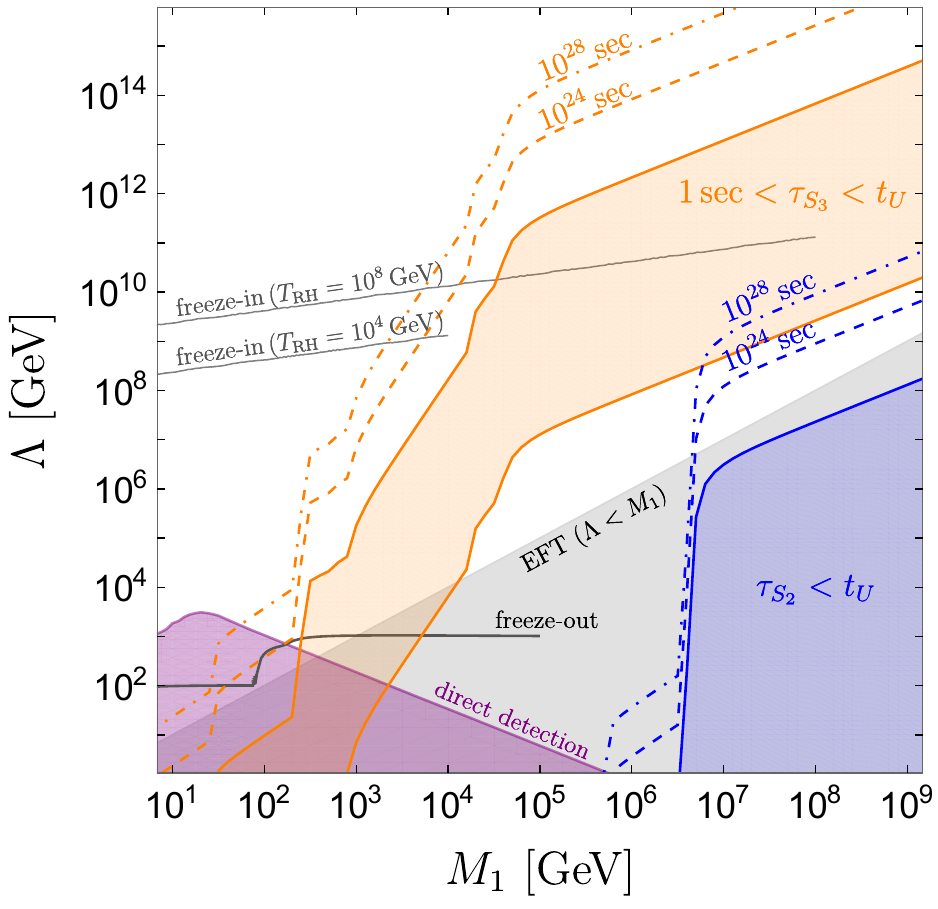}
    \includegraphics[width=0.48\textwidth]{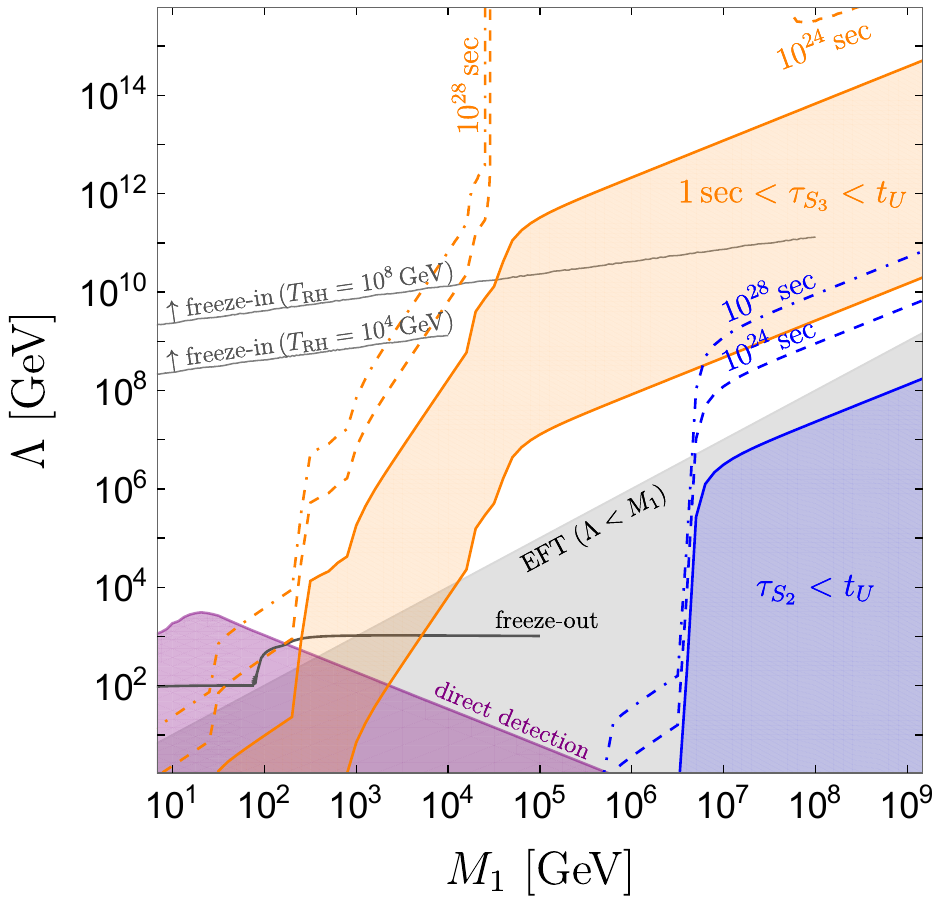}
    \caption{
    The constraints from the heavy scalar lifetimes for $\lambda=0$ (left) and $\lambda=10^{-11}$ (right). 
    The MFV expansion parameter is fixed to $\eps=10^{-2}$, which is related to the mass splitting as $\eps \simeq 2\Delta M/(y_t^2 M_1)\simeq 2\delta M/(y_c^2M_1)$. 
    The orange region, where $1\,{\rm sec} \leq \tau_{S_3} \leq t_U$, 
    is excluded from the $S_3$ stability and the BBN bound.
    In the blue region, $S_2$ is unstable and cannot be DM. 
    The blue (orange) dashed and dot-dashed lines respectively stand for contours of $\tau_{S_2} \, (\tau_{S_3}) = 10^{24}\sec$ and $10^{28}\sec$. 
    In the gray shaded region, we have $\Lambda \leq M_1$ and the EFT description is not justified. 
    On the black (gray) line, the DM abundance is correctly produced by the freeze-out (freeze-in) mechanism. 
    }
    \label{fig:lifetime_bound1}
\end{figure}

\begin{figure}[t]
    \centering
    \includegraphics[width=1\textwidth]{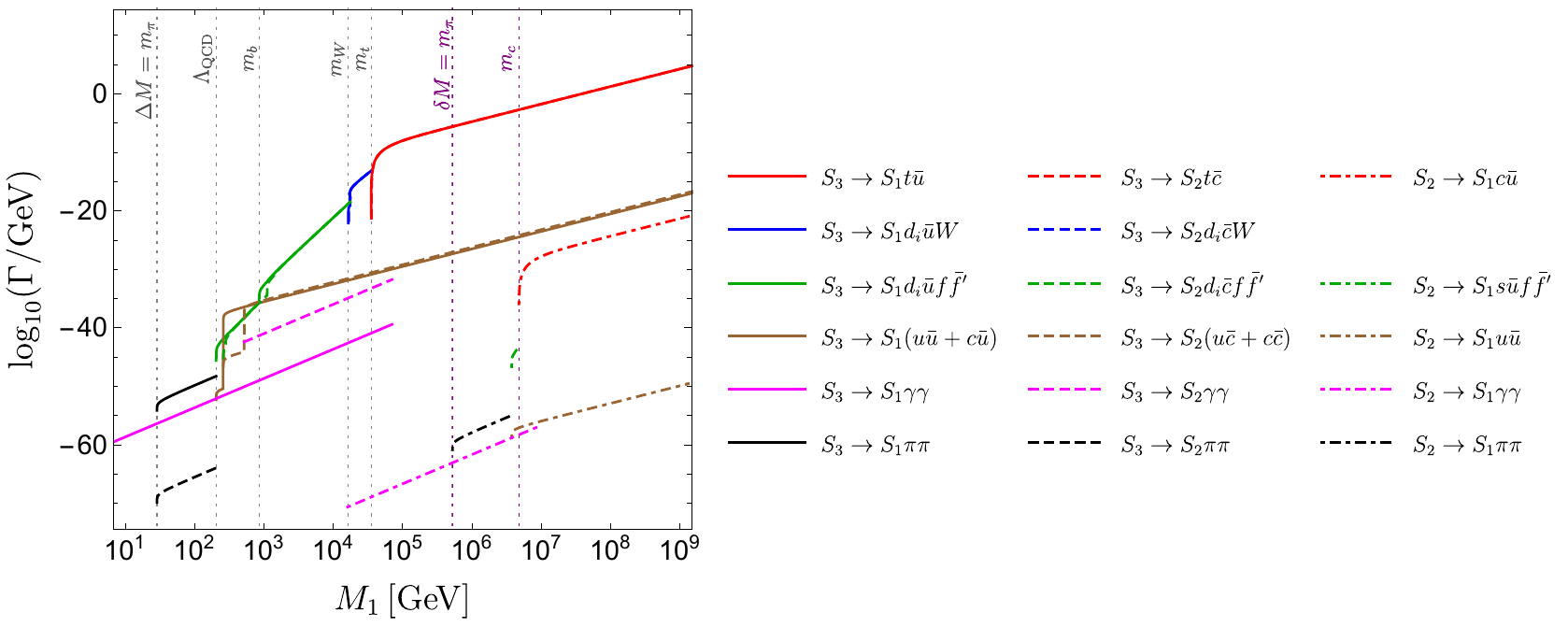}
    \caption{
    Partial decay widths of the heavy scalars, $S_2$ and $S_3$, induced from the dim-6 interactions. 
    We take $\Lambda=10^3\,\GeV$ and $\eps=10^{-2}$. 
    The MFV expansion parameter $\eps$ is related to the mass splitting as $\eps \simeq 2\Delta M/(y_t^2 M_1) \simeq 2\delta M/(y_c^2 M_1)$, where $\Delta M=M_3-M_1$ and $\delta M=M_2-M_1$. 
    The solid lines indicate the decay widths of $S_3$ into $S_1$, while the dashed lines represent those into $S_2$. The dot-dashed lines show the decay widths of $S_2$ into $S_1$.
    The vertical gray (purple) dotted lines show the representative kinematical thresholds for the $S_3$ ($S_2$) decay processes.
    }
    \label{fig:widths}
\end{figure}

In Fig.~\ref{fig:lifetime_bound1}, 
we show constraints from lifetimes of the heavy scalars $S_{2,3}$ with $\lambda=0$ (left) and $\lambda=10^{-11}$ (right). 
In the blue region, the lifetime of $S_2$ is shorter than the age of the universe $t_U$, while in the orange region the lifetime of $S_3$ lies in $1\,\sec < \tau_{S_3} < t_U$. 
Thus, two-component DM consisting of $S_1$ and $S_2$ is realized between the upper boundary of the blue region and the lower boundary of the orange region. 
Above the upper boundary of the orange region, all three scalars are stable and three-component DM scenario is realized.
The blue (orange) dashed and dot-dashed lines respectively stand for contours of $\tau_{S_2} \, (\tau_{S_3}) = 10^{24}\sec$ and $10^{28}\sec$, 
as a reference of constraints from indirect DM searches and cosmological observations. 
See discussion in Section \ref{sec:discussion} for further details. 
On the right panel, we set the Higgs portal coupling to $\lambda=10^{-11}$. 
Since it is very weak, the exclusion regions, filled with colors, are the same in two plots. 
The only difference is seen in the lifetime contours of $10^{24}\sec$ and $10^{28}\sec$ for high cutoff scales $\Lambda$.

The lifetime bounds in Fig.~\ref{fig:lifetime_bound1} exhibit several kinks, which appear at kinematical thresholds of decay processes. 
Figure \ref{fig:widths} shows partial decay widths for two heavy scalars, $S_2$ and $S_3$, that are induced solely by the dim-6 interactions. 
We take $\Lambda=10^3\,\GeV$ and $\eps=10^{-2}$ there. 
The widths for different $\Lambda$ are obtained by an overall scaling $\Gamma\propto 1/\Lambda^4$. 
For $S_3$ decay, each threshold appears when the mass splitting $\Delta M$ is around 
the top quark mass (at $M_1 \simeq 35\,\TeV$), 
the $W$ boson mass (at $M_1 \simeq 16\,\TeV$), 
the bottom quark mass (at $M_1 \simeq 860\,\GeV$), 
the charm quark mass (at $M_1 \simeq 260\,\GeV$), 
the QCD scale (at $M_1 \simeq 200\,\GeV$), which we take 1\,GeV, 
and the pion mass (at $M_1 \simeq 28\,\GeV$). 
The corresponding thresholds are shown by the vertical gray dotted lines. 
The processes induced at the higher order of $\eps$ surpass the leading-order five-body processes only below the bottom threshold. 
For $S_2$ decay, kinks are visible at $M_1 \sim 5\,{\rm PeV}$ and $0.5\,{\rm PeV}$ in Fig.~\ref{fig:lifetime_bound1}, which correspond to the charm quark and pion thresholds, respectively. 
See the vertical purple dotted lines in Fig.~\ref{fig:widths}. 
We add that our width calculation suffers from a large hadronic uncertainty around $\Delta M,\,\delta M \sim 1\,\GeV$, because of a complication of QCD dynamics. 
A special care to evaluate hadronic contributions is necessary in that region.

We also show other theoretical and experimental constraints in Fig.~\ref{fig:lifetime_bound1}. 
In the gray region, we have $\Lambda < M_1$ and the EFT description is not justified. 
The purple region is excluded by direct DM detection bound, see Appendix \ref{app:DirectDet} for the detail. 
On the black and gray lines, total relic abundance of stable flavored scalars can account for the observed DM abundance, $\Omega h^2 = 0.12$. 
We consider two production mechanisms:
one is the conventional thermal freeze-out production in the radiation dominated universe, and the other is the freeze-in production (see Appendix \ref{app:DMprod}). 
In the freeze-out case, only the dim-6 interactions are responsible for the production, 
since the Higgs portal coupling is vanishing or too weak to contribute. 
The observed DM abundance is explained for $\Lambda\simeq10^2$--$10^3\,\GeV$, although only a limited mass range $M_1 \simeq 180$--$210\,\GeV$ is compatible with the bounds from the $S_3$ lifetime and direct detection. 
Note that the freeze-out production does not work for $M_1 \gtrsim 100\,\TeV$ due to unitarity limit~\cite{Griest:1989wd, Smirnov:2019ngs}, so we simply cut off the black line at $M_1=100\,\TeV$. 
In the freeze-in case, the DM production crucially depends on the Higgs portal coupling. 
If that coupling is much weaker than $\lambda=10^{-11}$, the freeze-in production proceeds mostly through the dim-6 interactions with negligible Higgs portal contribution. 
The correct abundance is accommodated in $\Lambda\simeq10^7$\,--\,$10^{10}\,\GeV$, see the gray lines in Fig.~\ref{fig:lifetime_bound1} (left). 
The production rate with the dim-6 interactions is larger at higher temperatures. 
The DM abundance is thus sensitive to how the universe is reheated after inflation.  
We assume instantaneous reheating at a temperature $T_{\rm RH}$ in our freeze-in calculation and integrate Boltzmann equations from $T=T_{\rm RH}$ to $T=T_0$ with zero initial DM abundance. 
In contrast, if the Higgs portal coupling $\lambda$ amounts to $10^{-11}$ or larger, 
it can significantly contribute to the freeze-in production via $h \to S_i S_i^*$ and $h h \to S_i S_i^*$. 
The required coupling for the correct abundance is 
$\lambda\simeq2.2\times10^{-11}$ for $m_h < M_i$ and $\lambda\simeq1.2\times10^{-11} \sqrt{\frac{\GeV}{M_i}}$ for $m_h \gg M_i$ in a pure Higgs portal DM case~\cite{Lebedev:2019ton}. 
In this regime, the production depends insensitively on the reheating temperature if $T_{\rm RH} \gg M_i$. 
Instead, one has to tame thermalization and overproduction via the dim-6 interactions. 
These restrictions are avoided above the gray lines for a given reheating temperature, see Fig.~\ref{fig:lifetime_bound1} (right). 
It would be worth mentioning that in our benchmark model, two-component parameter spaces are not consistent with either the standard freeze-out or freeze-in production. 
Other production mechanisms or non-standard cosmological history should be considered there.

\begin{figure}[t]
\centering
    \includegraphics[width=0.48\textwidth]{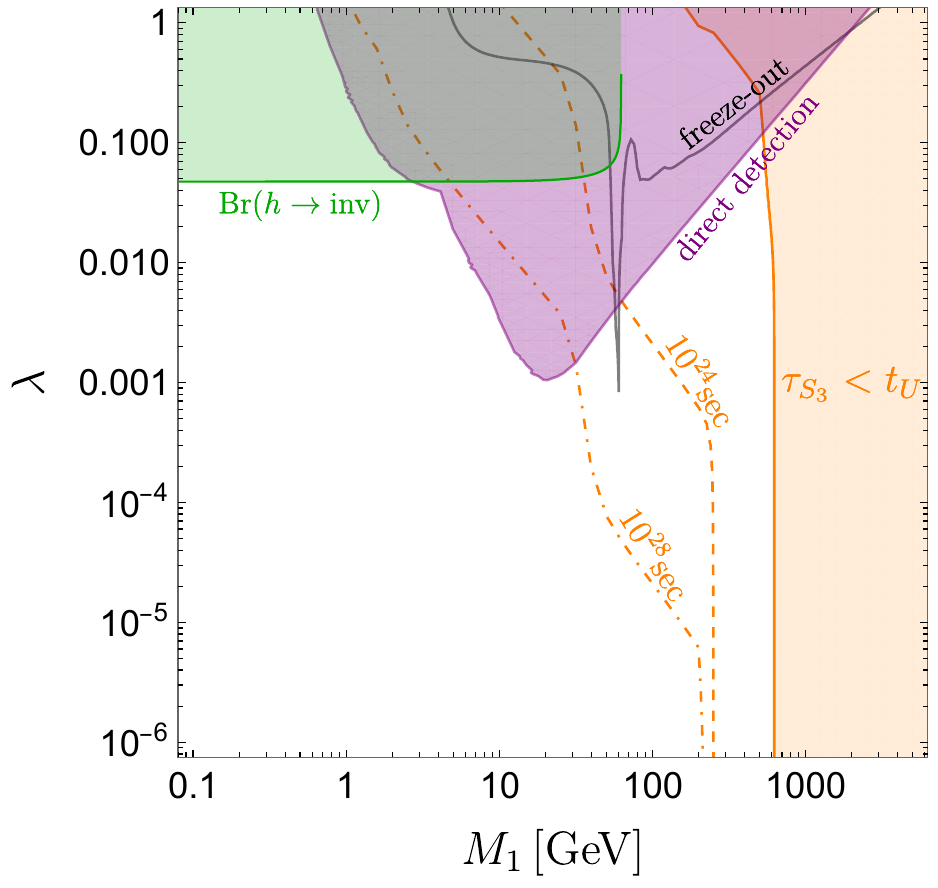}
    \includegraphics[width=0.48\textwidth]{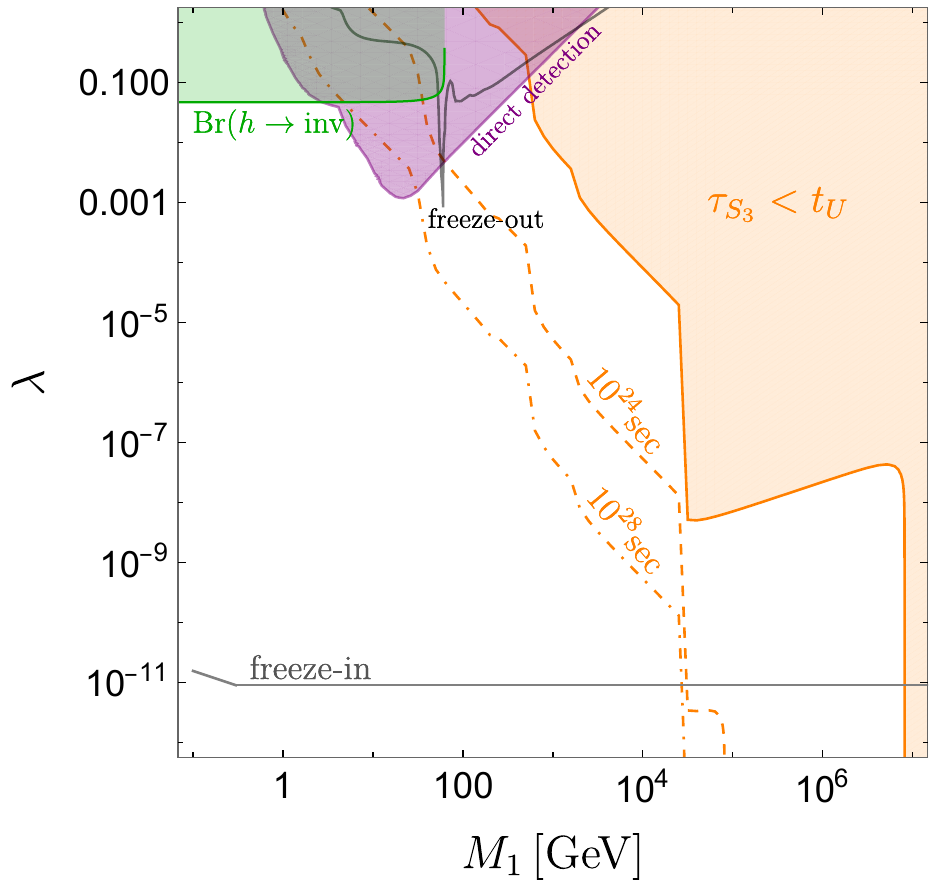}
    \caption{
    The lifetime constraints and experimental bounds for $\Lambda=10^4\,\GeV$ (left) and $\Lambda=10^{13}\,\GeV$ (right). 
    The MFV expansion parameter is fixed to $\eps=10^{-2}$, which is related to the mass splitting as $\eps \simeq 2\Delta M/(y_t^2 M_1)$. 
    Color coding of each constraint is the same as in Fig.~\ref{fig:lifetime_bound1}, except for the green region which is excluded by the Higgs invisible decay bound. 
    The lifetime of $S_2$ is longer than the age of the universe in the entire region. 
    The orange dashed and dot-dashed lines correspond to the lifetime contours of $\tau_{S_3}=10^{24}\sec$ and $10^{28}\sec$, respectively. 
    On the black (gray) line, the DM abundance is correctly produced by the freeze-out (freeze-in) mechanism.
    }
    \label{fig:lifetime_bound2}
\end{figure}

In Fig.~\ref{fig:lifetime_bound2}, 
we show the lifetime constraints in the $(M_1, \lambda)$ plane. 
The cutoff scale is fixed to $\Lambda=10^4\,\GeV$ (left) and $\Lambda=10^{13}\,\GeV$ (right). 
Color coding of each constraint is the same as in Fig.~\ref{fig:lifetime_bound1}, except for the green region which is excluded by the Higgs invisible decay bounds \cite{CMS:2018yfx, ATLAS:2020cjb}. 
On the black line, the DM abundance is correctly produced by the freeze-out mechanism with the Higgs portal coupling, albeit only in the Higgs resonance region $M_1 \simeq m_h/2$. 
The dim-6 interactions do not make significant contribution to the freeze-out. 
For $\Lambda=10^{13}\,\GeV$, the flavored scalars are not thermalized via the dim-6 interactions unless the reheating temperature is extremely high. 
In this case, the freeze-in production via the Higgs portal processes succeeds for $\lambda\sim 10^{-11}$~\cite{Lebedev:2019ton}.

\section{Discussion}
\label{sec:discussion}

We saw that within the MFV framework, DM can comprise multiple components that originate in one flavor multiplet. 
More than one component of those flavored states has sufficient longevity to serve as DM across a broad parameter space, while we have not studied their detailed phenomenology. 
In this section, we enumerate some brief comments on phenomenological implications of multi-component flavored DM, which are left for future works. 

\begin{itemize}

    \item {\bf Indirect searches for heavy decaying DM components}: 
    If heavy flavor components are DM, their present-time decay in galaxies produces a large number of energetic photons, positrons and neutrinos, which contribute to photon and cosmic-ray fluxes in space. 
    These additional fluxes are constrained by astrophysical observations of 
    gamma-rays~\cite{Cirelli:2012ut, Massari:2015xea, Cohen:2016uyg, Chianese:2019kyl, Esmaili:2021yaw, Maity:2021umk, Chianese:2021jke, Song:2023xdk, Dutta:2022wuc, Fermi-LAT:2012pls, VERITAS:2012cdo, Aleksic:2013xea, MAGIC:2018tuz, HAWC:2017mfa, HAWC:2017udy, HAWC:2023owv, HAWC:2023bti, LHAASO:2022yxw, Fermi-LAT:2015kyq, Foster:2022nva, DeLaTorreLuque:2023fyg}, 
    $X$-rays~\cite{Essig:2013goa, Boddy:2015efa, Ng:2019gch, Laha:2020ivk, Cirelli:2020bpc, Roach:2022lgo, Calore:2022pks, Cirelli:2023tnx, DelaTorreLuque:2023olp}, 
    radio-waves~\cite{Regis:2021glv, Dutta:2022wuc}, 
    positrons~\cite{Boudaud:2016mos,DelaTorreLuque:2023olp} and 
    neutrinos~\cite{IceCube:2011kcp, IceCube:2018tkk, IceCube:2023ies, Arguelles:2022nbl, Akita:2023qiz}. 
    See also a comprehensive review \cite{Cirelli:2024ssz} and references therein. 
    For a DM particle decaying only into SM particles, the current best lower limits on DM lifetimes reach $\tau_{\rm DM} \sim 10^{24}$--$10^{28}$\,sec~\cite{Cirelli:2024ssz}, depending on mass range and decay modes of DM. 
    These astrophysical bounds suggest that lifetimes of decaying DM have to be much longer than the age of the universe and, therefore, some of multi-component parameter spaces (figures \ref{fig:lifetime_bound1} and \ref{fig:lifetime_bound2}) might be excluded by comparing with photon and cosmic-ray observations. 

    \item {\bf Cosmological bounds}: Decay of heavy states into SM particles in the early universe leaves observable imprints on cosmology, even if their lifetimes are longer than the age of the universe. 
    For instance, exotic energy injection due to DM decay into SM particles has a significant impact on the ionization and thermal history of the universe, and distorts anisotropy spectra of Cosmic Microwave Background (CMB) \cite{Silk:1983hj, Kawasaki:1985ff, Hu:1993gc, Adams:1998nr, Chen:2003gz, Padmanabhan:2005es, Kanzaki:2007pd, Chluba:2011hw, Diamanti:2013bia, Liu:2016cnk, Slatyer:2016qyl}. 
    The CMB data currently impose lower bounds on lifetimes, $\tau_\DM \gtrsim 10^{24}$--$10^{25}\,\sec$~\cite{Liu:2016cnk, Slatyer:2016qyl, Capozzi:2023xie, Liu:2023nct}, 
    which are comparable with constraints from indirect DM searches. 
    There are also relevant constraints from 
    Lyman-$\alpha$~\cite{Liu:2020wqz, Capozzi:2023xie},
    21-cm~\cite{Shchekinov:2006eb, Furlanetto:2006wp, Valdes:2007cu, Cumberbatch:2008rh, DAmico:2018sxd, Clark:2018ghm, Cheung:2018vww, Liu:2018uzy, Mitridate:2018iag, Facchinetti:2023slb, Sun:2023acy} and  
    heating of a gas-rich dwarf galaxy Leo T~\cite{Wadekar:2021qae}. 
    While the current best cosmological bounds ($\tau_\DM \gtrsim 10^{25}\,\sec$) are derived from the CMB and Lyman-$\alpha$, it is remarkable that future HERA measurements of the 21-cm power spectrum can surpass the CMB and Lyman-$\alpha$ sensitivity and reach lifetimes of $10^{27}$--$10^{28}$\,sec~\cite{Facchinetti:2023slb, Sun:2023acy}. 
    That encourages us to pursue cosmological searches in addition to indirect DM searches. 
    
    \item {\bf Flavor physics}: 
    Although the quark flavor symmetry is naturally conserving in this framework, the intrinsic flavor violation from the CKM matrix can still accommodate additional contribution to flavor violating observables on the top of the SM contribution \cite{Batell:2011tc, DAmbrosio:2002vsn}. 
    The new physics effects can be analyzed on a model-by-model basis or in a general way by matching with the Standard Model Effective Field Theory (SMEFT) \cite {Buchmuller:1985jz, Grzadkowski:2010es, Leung:1984ni} with MFV Wilson coefficients \cite{Alonso:2013hga, Brivio:2017btx, Silvestrini:2018dos, Hurth:2019ula, Aoude:2020dwv, Faroughy:2020ina, Greljo:2022cah, Bruggisser:2022rhb, Greljo:2023adz, Bartocci:2023nvp, Greljo:2023bdy} if the new physics scales, $\Lambda$ or $M_i$, are high enough. 
    Besides, one potentially interesting phenomenon might be an apparent flavor violation from a natural-flavor-conserving new physics sector, which can occur due to the fact that DM particles carry quark flavor charges. 
    Such a process might have some implication for a recent Belle-II excess in the $B \to K \nu \nu$ process \cite{Belle-II:2023esi}. 
    As discussed in earlier works \cite{Fridell:2023ssf, Bolton:2024egx}, 
    a new three-body decay channel $B \to K \chi \chi$ with $\chi$ being an invisible particle provides a good fit to that excess (see also \cite{Bause:2023mfe, Buras:2024ewl, Felkl:2023ayn, He:2023bnk}). 
    This three-body process naturally appears in our framework via $b \to s \chi_3 \bar{\chi}_2$. 

    \item {\bf Inelastic scattering}: 
    Multiple states of DM with a small mass splitting leaves a unique signal at DM direct detection experiments through inelastic scattering, e.g. $\chi_i N \to \chi_j N$ \cite{XENON10:2009sho, ZEPLIN-III:2010cnv, CDMS-II:2010wvq, XENON100:2011hxw, Bramante:2016rdh, PandaX-II:2017zex, PandaX:2022djq}. 
    In general, both up-scattering and down-scattering off a nucleus are possible. 
    Such processes are known in the context of 
    (endothermic) inelastic DM ($M_i < M_j$) \cite{Tucker-Smith:2001myb, Tucker-Smith:2004mxa, Chang:2008gd} and 
    of exothermic DM ($M_i > M_j$) \cite{Finkbeiner:2009mi, Essig:2010ye, Graham:2010ca}. 
    The MFV framework would offer natural UV prescriptions for those inelastic DM scenarios. 
    
    \item {\bf Detection of boosted lighter components}: 
    Annihilation or decay of heavier components can produce lighter DM components with a velocity larger than their virial velocities in halos. 
    Such boosted DM components are detected at terrestrial experiments. 
    
\end{itemize}

In addition to the above-mentioned subjects, 
one can pursue model building of flavored DM in the MFV framework. 
In this paper, we only considered a gauge singlet scalar DM, while the stability discussion in Section \ref{sec:MFVDM} is applied for any spin and EW representations. 
Different choices of those representations, such as fermionic fields or EW multiplets, would result in different phenomenology. 
Additionally, one can include other new particles that reside around DM mass scales and mediate interactions between flavored DM and the SM fields. 
Such an extension would not spoil the DM stability unless the MFV ansatz and the flavor triality condition are violated. 
It could expand viable parameter spaces, making it compatible with the freeze-out and freeze-in production. 
Two-component DM parameter spaces might be enabled by this extension.

As a final remark, it should be noticed that the flavor trialilty condition is a sufficient condition for the DM stability, but not necessary. 
Thus, it is possible in general that one finds a specific combination of flavor, EW and Lorentz representations that does not satisfy the triality condition, but leads to an accidentally long-lived or absolutely stable neutral state. 
Such a new candidate could be systematically explored by employing the Hilbert series \cite{Grinstein:2023njq}.

\section{Summary}
\label{sec:summary}

The MFV hypothesis provides a robust framework for studying new physics models that include additional sources of flavor violation. 
As application of this framework to DM, it is established that the lightest component of a new flavored field can be naturally stabilized~\cite{Batell:2011tc}. 

In this paper, we investigated a possibility that, under the MFV framework, the heavy components of such a new flavored field are also stable over cosmological timescales and constitute a significant portion of DM. 
For illustration, we consider a gauge singlet, $\SU(3)_{u_R}$ triplet scalar field, which is one of the simplest candidates for flavored DM. 
All relevant interactions and the mass spectrum of the flavored scalars are governed by the quark Yukawa couplings, the CKM matrix and the MFV expansion parameter $\eps$, up to UV-model dependent ${\cal O}(1)$ coefficients. 
We evaluate the lifetimes of the heavy components as they decay into the lighter ones and SM particles. 
The decay processes are driven by the couplings to the Higgs boson and the dim-6 operators. 
The Higgs-mediated decay does not occur at the leading order of the MFV expansion and is significantly suppressed by the small expansion parameter $\eps$ and the light-quark Yukawa couplings. 
Conversely, the dim-6 operators induce the heavy scalar decay even in the $\eps\to0$ limit. 
Meanwhile, such decay processes are suppressed by the cutoff scale $\Lambda$, which can be extremely high. 

We identified parameter spaces where two or three components of the flavored scalar field have sufficient longevity to serve as DM. 
In the analysis, we adopt $\eps = 10^{-2}$, which is a minimum value induced from radiative corrections through the weak interactions. 
The parameter spaces for multi-component DM are derived by requiring that the lifetimes of the heavy states are longer than the age of the universe. 
These parameter spaces are compatible with the DM production in the conventional freeze-out and freeze-in mechanisms and the current direct detection bounds. 
See Figs.~\ref{fig:lifetime_bound1} and \ref{fig:lifetime_bound2} for our main results. 
In conclusion, multi-component flavored DM we proposed in this paper would provide a rich phenomenology and cosmology. 
Several implications are briefly mentioned in Section \ref{sec:discussion}. 
These subjects will be addressed in future.

\section*{Acknowledgements}
S.O. would like to thank Tomohiro Abe, Motoi Endo, Syuhei Iguro, Giacomo Landini, Luca Di Luzio, Seodong Shin and Pablo Qu\'ilez for fruitful discussions. 
The work of S.O. was supported by JSPS KAKENHI Grant Numbers JP22K21350 and by a Maria Zambrano fellowship financed by the State Agency for Research of the Spanish Ministry of Science and Innovation.
F.M. is supported in part by the INFN “Iniziativa Specifica” Theoretical Astroparticle Physics (TAsP). 
K.W. is supported also by the China Scholarship Council (CSC). 
This work also receives support from grants PID2022-126224NB-C21, PID2019-105614GB-C21 and Excellence Maria de Maeztu 2020-2023” award to the
ICC-UB CEX2019-000918-M, as well as from grants  2021-SGR-249 and 2017-SGR-929  (Generalitat de
Catalunya).

\appendix

\section{$N$-body phase space with cluster decomposition}
\label{sec:PhiN}

\subsection{Methodology}

Lorentz-invariant $N$-body phase space with invariant mass $M$ is defined by 
\begin{align}
\label{eq:def-PhiN}
\Phi_N(M^2; m_i^2) 
	& = \int \delta^4(Q-\sum_i^N p_i) \prod_i^N \frac{\d^3p_i}{(2\pi)^3 2E_i}  \,.
\end{align}
Here, we define $Q^2=M^2$ and $p_i^2=m_i^2$. 
The phase space $\Phi_N$ is a function only of $M^2$ and $m_i^2$ and useful to evaluate partial width for an $N$-body decay process $A(Q) \to \sum_i^N a_i (p_i)$,
\beq
\Gamma = \int \frac{(2\pi)^4}{2M} |\M|^2 \d\Phi_N \,.
\eeq

Using two-cluster decomposition~\cite{Block:1979sy}, 
\eq{eq:def-PhiN} can be decomposed into 
two clusters of $m$ and $n$ particles with $N=m+n$ 
(i.e. one being a cluster of the $m$ particles with invariant mass $M_1$
and the other a cluster for the remaining $n$ particles with invariant mass $M_2$), 
\beq
\Phi_N(M^2; \mu_i^2, \mu_j^2) 
	= \frac{\pi}{2} \int \d M_1^2 \, \d M_2^2 \,
	\Phi_m(M_1^2; \mu_i^2) F_1(M_1^2/M^2, M_2^2/M^2) \Phi_n(M_2^2; \mu_j^2) \,,
\eeq
where $\mu_i$ denote masses of the particles in the $m$-cluster
and $\mu_j$ in the $n$-cluster, 
and 
\beq
F_1(x,y) = \sqrt{1-2(x+y)+(x-y)^2} \,.
\eeq
In some case, it is convenient to introduce normalized masses, 
\begin{align}
x & = \frac{M_1^2}{M^2} , \quad u_i = \frac{\mu_i^2}{M^2} \,,\\
y & = \frac{M_2^2}{M^2} , \quad v_j = \frac{\mu_j^2}{M^2} \,.
\end{align}
and normalized phase space, 
\beq
\Phi_N(1; u_i, v_j) 
	= \frac{\pi}{2} \int \d x \, \d y \, \Phi_m(x; u_i) F_1(x,y) \Phi_n(y; v_j) \,,
\label{eq:PhiN-xy}
\eeq
which is related to \eq{eq:def-PhiN} as
\beq
\Phi_N(1; u_i, v_j) 
	= \frac{\Phi_N(M^2; \mu_i^2, \mu_j^2)}{(M^2)^{N-2}} \,.
\eeq
The integrand of Eq.~(\ref{eq:PhiN-xy}) can be understood as a joint distribution in $x,y$. 
Note that once applying \eq{eq:PhiN-xy} 
for $m=1$ (with its normalized mass being $x$) and $N=n+1$ clusters, 
we obtain 
\beq
\Phi_{n+1}(1; x, v_j) 
	= \frac{1}{(2\pi)^3} \frac{\pi}{2} 
		\int \d y \, F_1(x,y) \Phi_n (y; v_j) \,,
\eeq
which leads to another integral form of $\Phi_N(1; u_i, v_j)$, 
\beq
\Phi_N(1; u_i, v_j) 
	= (2\pi)^3 \int \d x \, \Phi_m(x; u_i) \Phi_{n+1}(1; x, v_j) \,.
\label{eq:PhiN-x}
\eeq
The latter expression is useful in some case.

For $N\geq4$, in general, 
the $N$-body phase space \eq{eq:def-PhiN} is not expressed in closed form 
except for two special situations: 
(i) all particles massless, 
(ii) one particle massive and the others massless. 
In addition, 
we have closed form phase space 
in a general case for $N=1,2$ and 
in a case with two particles massive and the other massless for $N=3$. 
We explicitly show those phase space expressions in the following subsections.

\subsection{For $N=1,2$}
The 1-body and 2-body phase spaces are trivial 
and found to be 
\begin{align}
\label{eq:Phi1}
\Phi_1 (M^2; m_1^2) 
	& = \frac{1}{(2\pi)^3} \, \delta(m_1^2-M^2) \,,\\
\label{eq:Phi2}
\Phi_2 (M^2; m_1^2, m_2^2) 
	& = \frac{1}{128\pi^5} \, F_1(m_1^2/M^2, m_2^2/M^2)\,.
\end{align}

\subsection{For $N=3$}
The 3-body phase space with two particles massive and one massless 
is given by 
\beq
\Phi_3(1; u_1, u_2, 0) 
	= 2 \Phi_3(1; 0) \int^1_{(\sqrt{u_1}+\sqrt{u_2})^2} \d y \, (1-y) F_1(u_1/y, u_2/y) \,,
\eeq
which is expressed explicitly in terms of the elementary functions, 
\begin{align}
\Phi_3(1; u_1, u_2, 0) 
	& = \Phi_3(1; 0) \, 
	\bigg\{ 
		(1+u_1+u_2) F_1(u_1, u_2) \nonumber\\
	&\quad + (u_1+u_2+|u_1-u_2|-u_1u_2) \ln(4u_1u_2) \nonumber\\
	&\quad + 2 \(2u_1u_2-u_1-u_2\) \ln|F_1(u_1,u_2)+1-u_1-u_2| \nonumber\\
	&\quad - 2 \left| u_1-u_2 \right|
			\ln\bigg|(u_1-u_2)^2-(u_1+u_2)+|u_1-u_2|F_1(u_1,u_2)\bigg| 
	\bigg\} \,.
\end{align}

\subsection{For any $N$ with all particles massless}

In a case with all particles massless, 
we have the $N$-body phase space in closed form for an arbitrary $N$. 
It is
\beq
\Phi_N(M^2; 0) = \frac{8(M^2)^{N-2}}{(4\pi)^{2N+1} (N-1)! (N-2)!} \,.
\label{eq:PhiN-M0}
\eeq
This is consistent with the result in \cite{DiLuzio:2015oha}.

We prove \eq{eq:PhiN-M0} here. 
To this end, we first relate $\Phi_N(1;0)$ to $\Phi_{N-1}(1;0)$.
Taking $m=N-1$ and $n=1$ in \eq{eq:PhiN-x} for all particles massless, 
$\Phi_N(1;0)$ is written by 
\begin{align}
\Phi_N(1;0) 
	& = (2\pi)^3 \int_0^1 \d x \, \Phi_{N-1}(x;0) \Phi_{2}(1; x, 0) \nonumber\\
	& = \frac{1}{(4\pi)^2} \int_0^1 \d x \, (1-x) \, \Phi_{N-1}(x;0) \nonumber\\
	& = \frac{1}{(4\pi)^2} \int_0^1 \d x \, (1-x) \, x^{(N-1)-2} \Phi_{N-1}(1;0) \nonumber\\
	& = \frac{\Phi_{N-1}(1;0)}{(4\pi)^2} \int_0^1 \d x \, (1-x) \, x^{N-3} \,,
\end{align}
where $\Phi_N(x;0)=x^{N-2}\Phi_N(1;0)$ is used in the third equality. 
Using the mathematical equality in the $\Gamma$ functions, 
\beq
\frac{\Gamma(x)\Gamma(y)}{\Gamma(x+y)} 
	= \int_0^1 \d t \, t^{x-1} (1-t)^{y-1} \,,
\eeq
$\Phi_N(1;0)$ is related to $\Phi_{N-1}(1;0)$ as 
\begin{align}
\Phi_N(1;0) 
	& = \frac{\Phi_{N-1}(1;0)}{(4\pi)^2} 
		\frac{\Gamma(N-2)\Gamma(2)}{\Gamma(N)} \nonumber\\
	& = \frac{\Phi_{N-1}(1;0)}{(4\pi)^2(N-1)(N-2)} \,.
\end{align}
Using this relation recursively, we get 
\begin{align}
\Phi_N(1;0) 
	& = \frac{1}{(4\pi)^2(N-1)(N-2)} 
	\times \cdots \times
	\frac{1}{(4\pi)^2(3-1)(3-2)} \Phi_2(1;0) \nonumber\\
	& = \frac{\Phi_2(1;0)}{(4\pi)^{2(N-2)} (N-1)! (N-2)!} \nonumber\\
	& = \frac{8}{(4\pi)^{2N+1} (N-1)! (N-2)!} \,.
\end{align}
In the end, \eq{eq:PhiN-M0} is easily obtained using $\Phi_N(M;0) = M^{2N-4} \, \Phi_N(1;0)$.

\subsection{For any $N$ with one particle massive and the others massless}

In a case with only one massive (its mass $\mu$) and the others massless, 
we also have a closed form phase space for any $N$. 
It is given by 
\beq
\Phi_N(1; v) 
    = \frac{8(N-1)(N-2)}{(4\pi)^{2N+1} (N-1)! (N-2)!}
	\int_0^{x_{\rm max}} \d x \, x^{N-3} F_1(x,v) \,,
\eeq
where $v=\mu^2/M^2$ and $x_{\rm max} = (1-\mu/M)^2$. 
It is easy to see that $\Phi_N(1;v)$ can be expressed in terms of 
$\Phi_N(1;0)$ (the $N$-body phase space with all massless particles),
\beq
\Phi_N(1;v) 
	= \Phi_N(1;0) f_{N-1}(v) \,,
\eeq
where we define $f_N(v)$ as 
\begin{align}
f_1(v) &:= F_1(v,0) = 1-v \,, \\
f_N(v) &:= N(N-1) \int_0^{x_{\rm max}} \d x \, x^{N-2} F_1(x,v) \,. \quad (N\geq2) \,
\end{align}
Below, we list explicit forms of $f_N(v)$ for $N=2,3,4$ for a practical purpose
\begin{align}
f_2(v) &= 1 - v^2 + 2 v \ln v \,,\\
f_3(v) &= 1 + 9v - 9v^2 - v^3 + 6v(1+v)\ln v \,,\\
f_4(v) &= 1 + 28v - 28v^3 - v^4 + 12v(1+3v+v^2)\ln v \,,
\end{align}
leading to 
\begin{align}
\Phi_2(1;v) &= \frac{1-v}{128\pi^5} \,,\quad\quad
\Phi_3(1;v) = \frac{f_2(v)}{4096\pi^7} \,,\\
\Phi_4(1;v) &= \frac{f_3(v)}{393216\pi^9} \,,\quad\quad
\Phi_5(1;v) = \frac{f_4(v)}{75497472\pi^{11}} \,.
\end{align}
One finds that $\Phi_2(1;v)$ above is consistent with \eq{eq:Phi2}. 
In some case, it is useful to expand $f_N(v)$ around $v=1$.
We have in the leading order
\begin{align}
f_2(v) \simeq \frac{1}{3} (1-v)^3 \,,\quad\quad
f_3(v) \simeq \frac{1}{10} (1-v)^5 \,,\quad\quad
f_4(v) \simeq \frac{1}{35} (1-v)^7 \,.
\end{align}
These expressions are used to evaluate the approximate decay widths in Section \ref{sec:decay}.

\section{Dark matter production}
\label{app:DMprod}

Regardless of whether DM is composed of a single component or multi-component, they have to be produced with the correct cosmological abundance in the early universe. 
In our model, DM can be produced from the SM plasma through the Higgs portal interactions and dim-6 operators. 
In this appendix, we evaluate DM relic abundance by taking the conventional thermal 
freeze-out \cite{Lee:1977ua, Hut:1977zn, Sato:1977ye, Vysotsky:1977pe} and 
freeze-in \cite{Ellis:1982yb, Nanopoulos:1983up, Ellis:1984eq, Berezinsky:1991kf, Moroi:1993mb, Kawasaki:1994af, Bolz:1998ek, Bolz:2000fu, McDonald:2001vt, Hall:2009bx, Elahi:2014fsa}\footnote{Freeze-in production is also discussed in the case of DM being 
axino~\cite{Covi:2001nw, Covi:2002vw}, sneutrino~\cite{Asaka:2005cn, Asaka:2006fs, Page:2007sh} and 
sterile neutrino~\cite{Shaposhnikov:2006xi, Kusenko:2006rh, Petraki:2007gq}.} 
as their production mechanisms. 
Other production mechanisms can succeed, depending on parameter choice and cosmological history. 

In both production mechanisms, the time evolution of number densities $n_i$ for $S_i$ is governed by Boltzmann equations, 
\beq
\frac{\d n_i}{\d t} + 3 H n_i 
    = 2 \int \frac{d^3{\bf p}_i}{(2\pi)^3 2E_i} \, {\cal C}[f_i] \,,
\label{eq:Boltzmann-eq}
\eeq
where $H$ is the expansion rate, 
\beq
H = \sqrt{\frac{8\pi G_N}{3} \rho} \,, \quad
\rho = \frac{\pi^2}{30} \, g_*(T) \, T^4 \,,
\eeq
with $T$ being the temperature of the SM plasma and $g_*(T) = g_{*,\,\SM}(T) + \sum_i g_{*,\,S_i}(T)$ the effective relativistic degrees of freedom. 
The collision term ${\cal C}[f_i]$ encodes all of DM number changing reactions induced from microphysical interactions. 
Focusing only on $2\to2$ processes, the pertinent contribution in our model comes from flavored scalar (co)annihilation $S_i S_j^* \to u_k \bar{u}_l$ and its inverse process, which are induced by the interactions in Eqs.~(\ref{eq:VS}) and (\ref{eq:O4-EWSB}). 
For simplicity, the Higgs portal interactions $\la_{hSi}$ are ignored here.\footnote{See \cite{Lopez-Honorez:2013wla} for the freeze-out production with the Higgs portal coupling in a single-component DM scenario.}
Then, the collision term takes the form, 
\begin{align}
{\cal C}[f_i] = 
    & - \frac{1}{2} \int 
        \frac{\d^3 {\bf p}_j}{(2\pi)^3 2E_j} 
        \frac{\d^3 {\bf p}_k}{(2\pi)^3 2E_k} 
        \frac{\d^3 {\bf p}_l}{(2\pi)^3 2E_l} 
        (2\pi)^4 \delta^{(4)}(p_i+p_j-p_k-p_l) 
        \left|\M(S_i S_j^* \to u_k \ol{u}_l)\right|^2 \nonumber\\
    & \quad \times 
        \left\{ 
            f_i({\bf p}_i) f_j({\bf p}_j) \[1-f_k({\bf p}_k)\] \[1-f_l({\bf p}_l)\] 
            - f_k({\bf p}_k) f_l({\bf p}_l) \[1+f_i({\bf p}_i)\] \[1+f_j({\bf p}_j)\] 
        \right\} \,,
\label{eq:collison_term}
\end{align}
where $f({\bf p})$ denotes momentum distribution for a particle species with four-momentum $p^\mu=(E, {\bf p})$ and $E=(m^2+{\bf p}^2)^{1/2}$, and we assume the time reversal is respected in the processes, i.e. 
$\left|\M(S_i S_j^* \to u_k \ol{u}_l)\right| = \left|\M(u_k \ol{u}_l \to S_i S_j^*)\right|$. 
The spin-summed squared amplitude is given by 
\begin{align}
\sum_{\rm spin} |{\cal M}(S_i S_j^* \to u_k \bar{u}_l)|^2
	& = \frac{N_c}{\Lambda^4} \bigg[ \(s-(m_u^k+m_u^l)^2\) \bigg\{ \(c_1\)^2 \((m_u^k)^2+(m_u^l)^2\) \delta_{ik} \delta_{jl} \nonumber\\
	& \quad + 2 c_1 c_2 \, m_u^k \(m_u^k+m_u^l\) \delta_{ij} \delta_{kl} \delta_{ik}
	+ 2 \(c_2\)^2 (m_u^k)^2 \delta_{ij} \delta_{kl} \bigg\} \nonumber\\
	& \quad + 2 \(c_1\)^2 m_km_l (m_k-m_l)^2 \delta_{ik} \delta_{jl} \bigg] \,.
	\label{eq:M2_SS-uubar}
\end{align}
In the freeze-out scenario, the annihilation evaluated at $s\simeq (M_i+M_j)^2$ determines the DM relic abundance, whereas the freeze-in production is most effective at $s \simeq T_\RH^2$, where $T_{\rm RH}$ is reheating temperature.

\subsection{Freeze-out}

In the freeze-out scenario, DM particles are assumed to be in thermal equilibrium with the SM plasma at high temperatures, when DM annihilation and creation reactions are balanced. 
As the universe expands and cools down, the rate of the reactions decreases and in the end, when the temperature cools down to $T\sim m_{\rm DM}/20$, the DM number changing processes are frozen and the DM abundance is fixed. 

Ignoring the quantum statistical factors and assuming that $S_i$ are in kinetic equilibrium with the thermal plasma, Eq.~(\ref{eq:collison_term}) is simplified to \cite{Gondolo:1990dk, Edsjo:1997bg}
\begin{align}
\frac{\d n_i}{\d t} + 3 H n_i 
	& = - \sum_{j,k,l} \vev{\sigma v_r}_{ij \to kl} \( n_i \, n_j - n_i^{\rm eq} \, n_j^{\rm eq} \) \,,
\label{eq:Boltzmann-FO}
\end{align}
where $n_i^{\rm eq}$ is the equilibrium number density of $S_i$, 
\beq
n_i^{\rm eq} (T) = \frac{m^2T}{2\pi^2} K_2(m/T) \,,
\eeq
with $K_2(x)$ the modified Bessel function of second kind of order 2.
Here, we assumed that DM is symmetric relic (i.e. $n_{S_i} = n_{S_i^*} = n_i$) and all SM particles follow the thermal distribution.
The thermal averaged cross section is defined by 
\beq
\vev{\sigma v_r}_{ij \to kl} 
    \equiv \frac{T}{32\pi^4} \frac{1}{n_i^{\rm eq}(T)\,n_j^{\rm eq}(T)} 
            \int_{s_{\rm min}}^\infty \d s \, \sigma_{ij \to kl} \, 
            \frac{\la(s,M_i^2,M_j^2)}{\sqrt{s}} \, K_1(\sqrt{s}/T) \,,
\label{eq:sigmavT}
\eeq
where $s_{\rm min}={\rm Max}\,[(M_i+M_j)^2,(m_u^k+m_u^l)^2]$ and $\sigma_{ij \to kl}$ is the cross section for the $S_i S_j^* \to u_k \bar{u}_l$ process. 

In parameter spaces where the freeze-out production succeeds, 
the cutoff scale will be around the EW scale. 
In order for the heavy flavored scalars to be stable and DM, they have to be highly degenerate with $S_1$. 
If unstable, instead, they have to decay into $S_1$ prior to the BBN.
In either case, the total energy density of DM is given to a good approximation by 
$\rho_\DM = 2 \sum_i M_i \, n_i \simeq M_1 \, n_\DM$, where $n_\DM = 2 \sum_i n_i$ and the prefactor of 2 counts DM and anti-DM. 
Moreover, conversion reactions, such as $S_1 t \leftrightarrow S_3 u$, occur more frequently than the annihilation reactions at the freeze-out time, because the number density of the SM particles is many orders of magnitude larger than that of the DM particles. 
Then, the fraction of the number densities of $S_i$ follows that of the equilibrium distributions, that is, 
\beq
\frac{n_i}{n_\DM} \simeq \frac{n_i^{\rm eq}}{n_\DM^{\rm eq}} \,.
\eeq
As a result, we obtain just a single equation for the time evolution of the total DM number density,
\begin{align}
\frac{\d n_\DM}{\d t} + 3 H n_\DM 
	& = - \vev{\sigma v_r}_\eff \[ \(n_\DM\)^2 - \(n_\DM^{\rm eq}\)^2 \] \,,
\end{align}
where the effective cross section is defined by 
\beq
\vev{\sigma v_r}_\eff = \sum_{i,j} \vev{\sigma v_r}_{ij} \frac{2 n_i^{\rm eq} \, n_j^{\rm eq}}{\(n_\DM^{\rm eq}\)^2} \,,
\eeq
with 
\beq
\vev{\sigma v_r}_{ij} = \sum_{k,l} \vev{\sigma v_r}_{ij \to k,l} \,.
\eeq
It is well known that $\vev{\sigma v_r}_\eff \simeq 3 \times 10^{-26} \cm^3/\sec$ at $T\simeq m_{\rm DM}/20$ provides the canonical cross section to produce the observed DM abundance in the freeze-out scenario. 

It is illuminating to estimate the thermal relic abundance in the case of $M_1 \simeq M_2 \simeq M_3$. 
In this case, the flavored scalars have comparable equilibrium densities, $n_1^{\rm eq} \simeq n_2^{\rm eq} \simeq n_3^{\rm eq} \simeq n_\DM^{\rm eq}/6$. 
Then, the effective cross section approximates to 
\beq
\vev{\sigma v_r}_\eff \simeq \frac{1}{18} \sum_{i,j} \vev{\sigma v_r}_{ij} \,.
\eeq
For $M_i \geq m_t$, the (co)annihilation processes involving top quarks in the final states dominate the production. 
The cross section for those processes in the non-relativistic limit $s\simeq (M_i+M_j)^2$ is given by 
\beq
\vev{\sigma v_r}_{ij} 
	\simeq \frac{N_c \, m_t^2}{4\pi \Lambda^4} \times 
		\left\{ \frac{(c_1)^2(\delta_{i3} + \delta_{j3})}{2} 
			+ 2 c_1 c_2 \delta_{i3} \delta_{j3}
			+ (c_2)^2 \delta_{ij} \right\} \,,
\eeq
leading to 
\beq
\vev{\sigma v_r}_\eff 
	\simeq \frac{N_c \, c^2 m_t^2}{9\pi \Lambda^4} 
	\simeq 3.3 \times 10^{-26} \cm^3/\sec \times \, 
                c^2 \(\frac{1\,\TeV}{\Lambda}\)^4 \,,
\eeq
with 
\beq
c^2 \equiv \frac{3 (c_1)^2 + 2 c_1 c_2 + 3 (c_2)^2}{8} \,.
\eeq
When the (co)annihilation into top quarks are kinematically forbidden, we need to take into account the (co)annihilation into lighter quarks. 
Taking $m_c \leq M_i \leq m_t/2$ for concreteness\footnote{This assumption is justified in the freeze-out case, since thermal relic DM with a lighter mass is excluded by the CMB measurements.}, we find the effective cross section to be
\beq
\vev{\sigma v_r}_\eff 
	\simeq \frac{N_c \, c^2 m_c^2}{9\pi \Lambda^4} 
	\simeq 2.0 \times 10^{-26} \cm^3/\sec \times c^2 \(\frac{100\,\GeV}{\Lambda}\)^4 \,.
\eeq
The cutoff scale takes $\Lambda \simeq 100\,\GeV\,\text{--}\, 1\,\TeV$ as anticipated.

\subsection{Freeze-in}

In the freeze-in scenario, 
it is assumed that DM particles never reach equilibrium with the SM plasma during the cosmological history. 
The time evolution of the flavored scalar number densities is governed by Boltzmann equations implementing only the one-way processes $u_k \bar{u}_l \to S_i S_j^*$. 
Assuming the Boltzmann distribution $f_{k,l} = e^{-E_{k,l}/T}$ for initial-state up-type quarks and ignoring unimportant quantum statistical factors for $S_i$, 
the collision integral takes the form, 
\begin{align}
{\cal N}(kl \to ij)
    \equiv \int \frac{d^3{\bf p}_i}{(2\pi)^3 E_i} \, {\cal C}[f_i] 
    = \frac{T}{32\pi^4} \int_{s_{\rm min}}^\infty \d s \, \sigma_{kl \to ij} \, \frac{\la(s,(m_u^k)^2,(m_u^l)^2)}{\sqrt{s}} \, K_1(\sqrt{s}/T) \,,
\end{align}
where $\sigma_{kl \to ij}$ denotes the cross section for $u_k \bar{u}_l \to S_i S_j^*$, 
\beq
\sigma_{kl \to ij} 
    = \frac{1}{4 \vmol E_k E_l} 
    \int \frac{d^3{\bf p}_i}{(2\pi)^3 2E_i} \frac{d^3{\bf p}_j}{(2\pi)^3 2E_j} 
    (2\pi)^4 \delta^{(4)}(p_k+p_l-p_i-p_j) |\M(u_k \bar{u}_l \to S_i S_j^*)|^2 \,,
\eeq
with $E_{k(l)}$ being the energy of $u_k (\bar{u}_l)$ in a reference frame, in which the M\o ller velocity is defined by 
\beq
\vmol = \sqrt{({\bf v}_k-{\bf v}_l)^2-({\bf v}_k\times{\bf v}_l)^2} \,.
\eeq
Compared with Eq.~(\ref{eq:sigmavT}), 
one realizes that the collision integral is expressed by the thermal averaged cross section, 
\beq
{\cal N}(kl \to ij) 
    = \vev{\sigma v_r}_{kl \to ij} n_k^{\rm eq}(T) \, n_l^{\rm eq}(T) \,.
\eeq
 
It is convenient to rewrite the Boltzmann equations using $Y_i = n_i/s$, where $s$ is the entropy density of the universe. 
Given $\frac{\d}{\d t} = - \widetilde{H} T \frac{\d}{\d T}$ in the radiation dominant universe, we find 
\begin{align}
\frac{\d Y_i}{\d T} 
    & = - \frac{1}{s\widetilde{H}T} \sum_{j,k,l} {\cal N}(kl \to ij) \,,
\label{eq:freezein-eq}
\end{align}
where $i=1,2,3$ and 
\beq
\widetilde{H} \equiv H \(1+\frac{1}{3} \frac{\d \ln g_{*s}}{\d \ln T} \)^{-1} \,.
\eeq
The present yield $Y_i(T_0)$ of each scalar is obtained by solving Eq.~(\ref{eq:freezein-eq}) from $T=T_\RH$ to $T=T_0$ with zero initial abundance $Y_i (T_\RH) = 0$ as a boundary condition. 
The total DM abundance is calculated by 
\beq
\Omega_\DM = \frac{s_0}{\rho_{\rm crit}} \times 2 \sum_i M_i \, Y_i (T_0) \,.
\label{eq:OmegaDM}
\eeq
With the values of the present entropy density $s_0$ and critical density $\rho_{\rm crit}$ \cite{ParticleDataGroup:2022pth},
\beq
s_0 = 2891.2 \, \cm^{-3} , \quad 
\rho_{\rm crit} = 1.053672(24)\times10^{-5}\, h^2 \, \GeV/\cm^3 \,,
\eeq
we obtain 
\beq
\Omega_\DM h^2 \simeq 0.12 \times \sum_i \frac{M_i}{100\,\GeV} \frac{Y_i (T_0)}{2.2\times10^{-12}} \,.
\label{eq:Omegah2}
\eeq

Let us estimate the total DM abundance for $M_i, \, m_u^i \ll T_\RH \leq \Lambda$. 
Assuming instantaneous reheating and $Y_i(T_\RH) \simeq 0$, the solution of the Boltzmann equations is given by 
\beq
Y_i (T_0) \simeq \int_{T_0}^{T_\RH} \d\ln{T} \, 
        \frac{1}{s\widetilde{H}} \sum_{k,l,j} {\cal N}(kl \to ij) \,.
\label{eq:Yi}
\eeq
The integrand is proportional to $T^1$, since ${\cal N}(kl \to ij) \propto T^6$, $s(T) \propto T^3$ and $\widetilde{H} \simeq H \propto T^2$ at high enough temperatures. 
This means that the DM production occurs most efficiently at $T \simeq T_{\rm RH}$, and Eq.~(\ref{eq:Yi}) approximates to 
\beq
Y_i (T_0) \simeq \gamma_i \, T_\RH \,,\quad\quad
\gamma_i \simeq \frac{1}{s H T} \sum_{k,l,j} {\cal N}(kl \to ij) \bigg|_{T=T_{\rm RH}} \,.
\label{eq:gammai}
\eeq
Given that at high temperatures the cross section approximates to
\beq
\sum_{k,l} \vev{\sigma v_r}_{kl \to ij} 
    \simeq \frac{N_c m_t^2}{8\pi \Lambda^4} \times 
	\left\{ \frac{(c_1)^2(\delta_{i3} + \delta_{j3})}{2} 
		+ 2 c_1 c_2 \delta_{i3} \delta_{j3}
		+ (c_2)^2 \delta_{ij} \right\} \,,
\eeq
we find 
\beq
\gamma_i \simeq 0.5 \times 10^{-14} \, c_i^2 \(\frac{10^{10}\,\GeV}{\Lambda}\)^4 \(\frac{1}{10^8\,\GeV}\) \,,
\label{eq:gammai_app}
\eeq
where we take $m_t=162\,\GeV$ and $g_*=g_{*,s}=106.75$ and 
\beq
c_i^2 := \sum_j \[ \frac{(c_1)^2(\delta_{i3} + \delta_{j3})}{2} 
		+ 2 c_1 c_2 \delta_{i3} \delta_{j3}
		+ (c_2)^2 \delta_{ij} \] \,.
\eeq
The freeze-in abundance is given by 
\begin{align}
\Omega_\DM h^2 
	& \simeq 0.12 \times \(\frac{10^{10}\,\GeV}{\Lambda}\)^4 
		\(\frac{T_\RH}{10^8\,\GeV}\) \sum_i  c_i^2 \(\frac{M_i}{45\,\TeV}\) \,.
\end{align}
We have confirmed that this abundance estimate agrees with numerical results calculated by {\tt micrOMEGAs\_5\_2\_4} \cite{Belanger:2018ccd} in an appropriate limit.

\section{Direct detection with nuclear recoils}
\label{app:DirectDet}

DM can be directly detected in terrestrial experiments through scattering off nucleons and electrons in a target material. 
This direct detection approach provides strong constraints on DM candidates produced by the thermal freeze-out mechanism. 

\begin{table}[t]
\begin{center}
\begin{tabular}{|c|c||c|c|}
        \hline 
        $f_{n,u}$ & 0.0110 & $f_{p,u}$ & 0.0153 \\ \hline  
        $f_{n,d}$ & 0.0273 &  $f_{p,d}$ & 0.0191 \\ \hline
        $f_{n,s}$ & 0.0447 &  $f_{p,s}$ & 0.0447 \\
    \hline 
\end{tabular}
    \caption{
    The nucleon matrix elements for the light quarks. 
    The values correspond to those of the {\tt micrOMEGAs} default \cite{Belanger:2008sj}.}
    \label{tab:fN}
\end{center}
\end{table} 

In our case, DM can (in)elastically scatter off nucleons in a target nucleus, $S_i N \to S_j N$, with the dim-6 operators and the Higgs portal interactions. 
For a while, we ignore the latter interactions. 
Since in the case of inelastic scattering, terrestrial direct detection experiments can only probe a small mass splitting $\Delta M_S \lesssim {\cal O}(100)\,\keV$ which is out of our scope, 
we focus on the elastic scattering case here.\footnote{A larger mass splitting up to $\Delta M_S \lesssim 100\,\MeV$ might be tested with future neutron star surface temperature observations \cite{Baryakhtar:2017dbj, Raj:2017wrv, Bell:2018pkk, Acevedo:2019agu, Fujiwara:2022uiq, Alvarez:2023fjj}, although recent studies discuss a possibility that built-in heating mechanisms of neutron stars would conceal extra heating through DM scattering and annihilation \cite{Yanagi:2019vrr, Hamaguchi:2019oev, Fujiwara:2023hlj, Fujiwara:2023tmr}.}
After the EW symmetry breaking and in the leading MFV expansion, 
the relevant interaction Lagrangian is given by Eq.~(\ref{eq:O4-EWSB}).
From these interactions, we find spin-independent cross section for $S_i$-nucleon elastic scattering, 
\beq
\sigma_{{\rm SI}, i} 
	= \frac{\mu^2 m_N^2}{4\pi M_i^2} \frac{1}{\Lambda^4} 
	\( \frac{Z C_{p,i} + (A-Z) C_{n,i} }{A} \)^2 \,.
\label{eq:sigmaSI}
\eeq
Here, $\mu = m_N M_i/(m_N+M_i)$ is DM-nucleon reduced mass and $Z$ and $A$ are atomic number and mass of a target nucleus, and 
\beq
C_{N, i} = c_1 \, f_{N, u_i} + c_2 \(f_{N,u}+f_{N,c}+f_{N,t}\), \quad \text{for $N=p,n$}
\eeq
with nucleon matrix elements $f_{N,q}$ for quarks $q$, 
\beq
\langle N| m_q \bar{q} q |N\rangle = m_N f_{N,q} \,.
\eeq
Using the QCD trace anomaly matching, the nucleon matrix elements for heavy quarks $Q$ are related to the one for gluon. 
At the leading order, we find 
\beq
\langle N| m_Q \bar{Q} Q |N\rangle 
    \simeq - \langle N| \frac{\alpha_s}{12\pi} G_{\mu\nu} G^{\mu\nu} |N\rangle 
    = \frac{2}{27} m_N f_{N,g} \,,
\eeq
where 
\beq
m_N f_{N,g} = - \langle N| \frac{9\alpha_s}{8\pi} G_{\mu\nu} G^{\mu\nu} |N\rangle \,
\eeq
and $f_{N,g} = 1 - \sum_{q=u,d,s} f_{N,q}$. 
See Table \ref{tab:fN} for the values of $f_{N,q}$ for light quarks, which we use in our numerical analysis. 
Note that effects of non-vanishing Higgs portal couplings can be easily included by 
modifying in Eq.~(\ref{eq:sigmaSI}) as 
\begin{align}
C_{N,i} & \to C_{N,i} + \la_{hSi} \frac{\Lambda^2}{m_h^2} \bigg(\frac{2}{9} + \frac{7}{9} \sum_{q=u,d,s} f_{N,q} \bigg) \,.
\end{align}

\begin{figure}[t]
\centering
\includegraphics[width=0.6\textwidth]{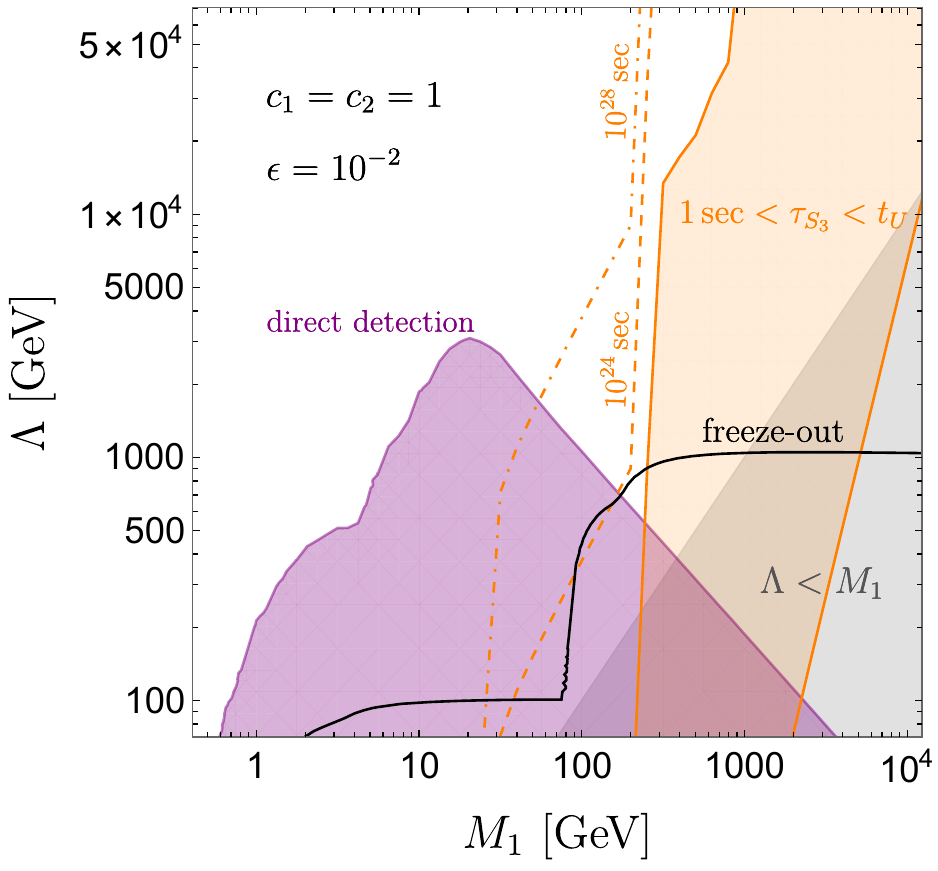}
    \caption{
    Direct detection bound (purple) with $c_1=c_2=1$. 
    The MFV expansion parameter is set to $\eps=10^{-2}$, which is related to the mass splitting as $\eps\simeq \Delta M/(y_t^2M_1)$. 
    The black line corresponds to the total abundance being equal to the observed value, 
    i.e. $\sum_{i=1}^3 \Omega_{S_i+S_i^*} h^2 = 0.12$, where we assume the freeze-out production. 
    The orange region is excluded by the constraint on the $S_3$ lifetime, where $1\,\sec < \tau_{S_3} < t_U$. 
    The orange dashed and dot-dashed lines correspond to contours of $\tau_{S_3} = 10^{24}\sec$ and $10^{28}\sec$.  
    }
    \label{fig:DirectDet}
\end{figure}

In Fig.~\ref{fig:DirectDet}, we show the current direct detection bound with $c_1=c_2=1$ and $\lambda=0$, which excludes the purple shaded region. 
The MFV expansion parameter is set to $\eps=10^{-2}$, which is related to the mass splitting as $\eps\simeq \Delta M/(y_t^2M_1)$, see also Eq.~(\ref{eq:DelM}). 
The black line corresponds to the total abundance being equal to the observed value, i.e. $\sum_{i=1}^3 \Omega_{S_i+S_i^*} h^2 = 0.12$, where we assume the freeze-out production. 
The orange region is excluded by the constraint on the $S_3$ lifetime, where $1\,\sec < \tau_{S_3} < t_U$. 
The orange dashed and dot-dashed lines correspond to contours of $\tau_{S_3} = 10^{24}\sec$ and $10^{28}\sec$. 
In the gray shaded region, we have $\Lambda \leq M_1$, 
where the EFT description is not justified.

\begin{small}
\bibliographystyle{utphys}
\bibliography{bibliography}
\end{small}

\end{document}